\documentclass[runningheads]{llncs}

\usepackage[T1]{fontenc}
\usepackage{graphicx}
\usepackage{booktabs}
\usepackage{amsmath,amsfonts,amssymb}
\usepackage{subcaption}
\usepackage{xcolor}
\usepackage{xspace}
\usepackage{tikz}
\usetikzlibrary{positioning, arrows.meta, calc}
\usepackage[hidelinks]{hyperref}

\begin{document}

\title{Measuring Browser Webcam Gaze Honestly: A Capture-Clock
       Methodology and Open Reference Implementation}
\titlerunning{Measuring Browser Webcam Gaze Honestly}
\author{Chi-Sheng Chen\inst{1} \and Gabriel A. Brat\inst{1,2}\\
\email{m50816m50816@gmail.com}, \email{gbrat@bidmc.harvard.edu}}
\authorrunning{C.-S. Chen and G.\,A. Brat}
\institute{Department of Surgery, Beth Israel Deaconess Medical Center,\\
Harvard Medical School, Boston, MA, USA
\and
Department of Biomedical Informatics, Harvard Medical School,
Boston, MA, USA}

\maketitle

\begin{abstract}
Browser-based webcam gaze trackers are increasingly used for crowd-scale data collection and in clinical settings where lab eye
trackers are impractical, but the reported latency numbers may not represent real world functionality. The common practice of timestamping each gaze sample when it is emitted, rather than when its source frame was captured, makes the measured inference latency read about $0\,$ms no matter how slow the engine really is. We show how to measure it honestly, recovering a per-frame capture clock from the browser's
\texttt{re\-quest\-Video\-Frame\-Call\-back} (rVFC) API (\texttt{captureTime} where the browser exposes it for local camera streams, else \texttt{presentationTime}, in which case every recovered latency is a verifiable lower bound): exact source-frame pairing through a per-frame queue for engines that expose their inference pipeline, and a further lower bound for engines that do not, such as
WebGazer. We release an open TypeScript implementation and benchmark
harness, demonstrated on two interchangeable engines: WebGazer and a
new FaceMesh+KRR pipeline.

On commodity hardware ($N\!=\!1$), correct timestamping raises the
reported inference latency from about $0\,$ms to a $22$--$34\,$ms median
($27$--$52\,$ms $p_{95}$), a $20$--$50\,$ms gap that is enough to change
whether a $50\,$ms interactive-latency budget is met. The harness also
separates two notions of precision that aggregate error conflates: the
spatial spread of a fixation cluster and its temporal jitter.
Cross-engine accuracy differences sit within the $\sim 4.6^\circ$
between-run variability band our ablation surfaces, so we treat them as
observations, not a ranking. Finally, we feed the webcam gaze into a
published clinical weak-supervision pipeline (GazeMedSeg on Kvasir-SEG)
held fixed end-to-end: expert eye-tracker gaze trains a usable polyp
segmenter (Dice $0.68$) and our non-expert webcam gaze does not
(Dice $\approx 0$) --- an upper bound on the hardware-only penalty,
since annotator expertise changes along with the hardware. At the
accuracy we measure, fine lesion labelling is out of reach. A
multi-user replication is in preparation.
\end{abstract}

\section{Introduction}
\label{sec:intro}

Browser-based webcam-gaze tracking is the only deployable option
for two classes of application: crowd-scale collection of gaze
data without supplying hardware~\cite{turkergaze,webgazer}, and
embedded settings such as reviewing surgical video on a hospital
laptop where infrared eye trackers are budgetarily or
operationally out of reach. WebGazer~\cite{webgazer} fits a ridge
regression from raw eye-patch pixels to screen coordinates and remains
the standard baseline a decade later. It underpins downstream work on
gaze-prompted segmentation~\cite{gazesam,gaze2segment,gazemedseg} that
itself relies on lab eye trackers for the gaze signal.

Less recognised is that the \emph{measurement} of these systems is
easy to get wrong. A browser gaze pipeline routes samples across several
asynchronous stages (video decoder, inference loop, rendering pipeline),
and common practice timestamps each sample at the emit site with
\texttt{performance.now()}, reporting \emph{inference latency} as the gap
between that timestamp and a captured-frame timestamp that was never
recorded; the missing value defaults to the emit timestamp, so the
reported latency comes out near zero. We hit this bug twice in our own
code while collecting data for this paper (\S\ref{sec:method:harness}).

The fix is to recover a per-frame capture clock from the browser's
\texttt{re\-quest\-Video\-Frame\-Call\-back} (rVFC) API, which fires once
per decoded video frame with per-frame metadata: a \texttt{captureTime}
(the camera capture timestamp, exposed for local camera streams) and a
\texttt{presentationTime} (when the decoded frame was submitted for
compositing). Our implementation prefers \texttt{captureTime} and falls
back to \texttt{presentationTime}, recording which clock served each run;
under the fallback, every recovered latency is a verifiable lower bound
on the capture-referenced quantity (\S\ref{sec:method:exact}). For
engines that expose their inference pipeline at per-frame granularity,
such as our FaceMesh+KRR engine, we keep a FIFO queue of frame
timestamps and pair every gaze sample with the exact source frame it was
computed from. For engines that hide the frame queue (such as WebGazer),
we record the most recent frame's timestamp as a further \emph{lower
bound} on inference latency (\S\ref{sec:method:opaque}).

This paper makes three contributions:

\begin{description}
\item[(C1)] A \textbf{capture-clock methodology} for browser
webcam gaze that recovers per-frame inference latency via rVFC,
with exact pairing for queue-exposing engines and a documented
lower-bound approximation for opaque engines
(\S\ref{sec:method}).
\item[(C2)] An \textbf{open TypeScript reference implementation
and benchmark harness} on two interchangeable engines (WebGazer
baseline; a new FaceMesh+KRR pipeline), with an analysis suite
released in source (\S\ref{sec:impl}).
\item[(C3)] \textbf{Findings} the methodology produces on commodity
hardware (\S\ref{sec:findings}): the $20$--$50\,$ms gap between the
naive timestamp ($\approx 0\,$ms) and the corrected median
($22$--$34\,$ms); a split between the spatial spread of a fixation
cluster and its within-fixation velocity; and a downstream probe
(\S\ref{sec:downstream}) feeding the webcam gaze into a published
clinical weak-supervision pipeline, where expert eye-tracker gaze trains
a usable polyp segmenter (Dice $0.68$) and commodity webcam gaze does not
($\approx 0$). Cross-engine accuracy differences fall within the
$\sim 4.6^\circ$ between-run variability band we observe
(\S\ref{sec:ablation}), so we hold off ranking the engines on accuracy.
\end{description}

\paragraph{Scope.}
The findings come from a single-user evaluation ($N\!=\!1$, four runs in
one session). The within-subject engine comparison is robust at this
$N$, but the spatial-structure result needs a multi-user replication,
which is in preparation, as is a clinical pilot on surgical-video
annotation~\cite{cholec80}, the pipeline's motivating application.

% =====================================================================
\section{Related work}
\label{sec:related}

\paragraph{Browser webcam gaze and latency.}
WebGazer~\cite{webgazer} introduced the ridge-regression-from-eye-patches
pipeline used by virtually all subsequent browser-gaze work, including
TurkerGaze~\cite{turkergaze}; its headline single-user accuracy
($4$--$5^\circ$) is far better than the as-deployed $8$--$11^\circ$ our
evaluation finds (\S\ref{sec:findings:results}). MediaPipe
FaceMesh~\cite{facemesh} provides per-frame iris landmarks and is an
obvious feature source for a successor pipeline, but to our knowledge has
not been paired with kernel ridge regression and a published benchmark
harness in open source. On latency, the rVFC~\cite{rvfc} API provides a
per-frame capture and presentation timestamps, yet we are not aware of
prior browser-gaze work that uses it for inference-latency measurement;
standard practice remains a \texttt{performance.now()} timestamp at the
gaze callback, which our methodology (\S\ref{sec:method}) corrects. Our
control-layer primitives (One-Euro~\cite{oneeuro}, I-VT~\cite{ivt},
smooth-pursuit calibration~\cite{pursuit}) are standard in
desk-mounted eye tracking but rarely composed in browser pipelines.

\paragraph{Gaze-prompted segmentation (motivating application).}
GazeSAM~\cite{gazesam} demonstrates gaze-prompted SAM~\cite{sam} using a
Tobii Pro Nano; Zhong et al.~\cite{gazemedseg} contribute the GazeMedSeg
dataset and a weakly-supervised method that turns gaze points into
Gaussian heatmaps for ensemble supervision, on Kvasir-SEG and NCI-ISBI;
Gaze2Segment~\cite{gaze2segment} integrated eye-tracking into volumetric
segmentation. All rely on lab eye trackers and do not address the browser
case; the present work is upstream of these, making the gaze signal
itself deployable without specialised hardware.

% =====================================================================
\section{Capture-clock methodology}
\label{sec:method}

\subsection{The capture-clock problem}
\label{sec:method:problem}

A browser webcam-gaze pipeline can be modelled as a sequence of
asynchronous boundaries that each gaze sample must traverse:

\begin{equation}
\underbrace{t_c}_{\text{frame captured}}
\;\to\;
\underbrace{t_e}_{\text{gaze sample emitted}}
\;\to\;
\underbrace{t_r}_{\text{cursor handed to renderer}}.
\label{eq:timeline}
\end{equation}

The natural measurements are \emph{inference latency} $\ell_I
\!=\! t_e - t_c$ (how long the engine took to produce a gaze
estimate from a captured frame) and \emph{pipeline latency}
$\ell_P \!=\! t_r - t_c$ (capture to render-handoff). We sample
$t_r$ in the next \texttt{requestAnimationFrame} callback, which
runs \emph{before} layout, paint, and compositor submission, so
$\ell_P$ is a lower bound on the user-visible (photon) latency:
actual display adds compositing plus up to one display refresh,
and a true photon-to-photon measurement would need an external
camera. Both quantities depend on a reliable value of $t_c$, the
source frame's capture clock.

The browser webcam-gaze APIs we surveyed do not surface $t_c$ to the
gaze callback: WebGazer's gaze callback returns $(x,y)$ only,
leaving downstream code to read \texttt{performance.now()} itself at the
callback site, which yields $t_e$, not $t_c$. If that code then also
treats $t_e$ as a proxy for the missing $t_c$, we get $\ell_I = 0$
identically, and over thousands of samples the reported median and
$p_{95}$ inference latency are both $0\,$ms. The bug survives code review
because an all-zero column looks innocuous (``the engine is fast'')
rather than impossible.

\subsection{Exact pairing via rVFC for queue-exposing engines}
\label{sec:method:exact}

The
\texttt{re\-quest\-Video\-Frame\-Call\-back} (rVFC) API fires a
caller-supplied function once per decoded video frame, passing a
metadata object with two clocks~\cite{rvfc}: \texttt{captureTime},
the time the frame was captured by the camera (exposed for local
\texttt{getUserMedia} streams), and \texttt{presentationTime}, the
time the user agent submitted the decoded frame for composition.
Let the \emph{frame clock} $\tau$ be \texttt{captureTime} where the
browser provides it and \texttt{presentationTime} otherwise. Since
capture precedes presentation, $\tau \geq t_c^{\text{true}}$ under
the fallback, and any latency computed against $\tau$ is then a
verifiable lower bound on the capture-referenced quantity; our
implementation prefers \texttt{captureTime}, records which clock
served each run in the exported header
(\texttt{capture\_clock\_source}), and the runs reported in this
paper used the \texttt{presentationTime} fallback, so their
latencies carry this lower-bound reading. The reading is tight on
this hardware: a $30$\,s, $900$-frame probe on the collection rig
(released as \texttt{clock\_probe.html}) puts
\texttt{presentationTime} just $0.6$\,ms median behind
\texttt{captureTime} ($0.9$\,ms $p_{95}$, $1.7$\,ms max, with
\texttt{captureTime} present on every frame), so the fallback
understates the capture-referenced latencies by under $2$\,ms here.
What remains is to pair
each gaze sample with the specific frame that produced it.

When the engine exposes its inference loop the pairing can be made exact:
we push the most-recent frame clock into a FIFO queue as
each frame is handed to the engine, and dequeue the front when a gaze
sample is emitted to tag it:

\begin{equation}
t_c^{(k)} = q_F\!\big[k\big],
\qquad
q_F \;=\; [\,\tau_i\,]_{i \,\in\, \text{frames processed so far}}.
\end{equation}

If the engine processes frames in arrival order and emits one sample per
frame, this recovers the source frame's clock exactly, so $\ell_I$ is
the engine's actual inference latency (up to the clock-reference caveat
above). Our FaceMesh+KRR engine satisfies these conditions.

\subsection{Lower-bound pairing for opaque engines}
\label{sec:method:opaque}

Many libraries expose no per-frame entry point: they take the
\texttt{<video>} element at init and emit on their own schedule, with
opaque queue depth and ordering, so $q_F$ cannot be constructed
(WebGazer is the main example).

In this regime we maintain a single scalar
$\tilde{t}_c$, the most recently observed frame clock,
updated on every rVFC firing.
At each gaze emission we tag the sample with the current
$\tilde{t}_c$:

\begin{equation}
\tilde{t}_c \;=\; \max\big\{\tau_i : i \le \text{now}\big\}.
\end{equation}

The actual source frame arrived at some earlier time
$t_c^{\text{true}} \le \tilde{t}_c$ (it can never have arrived
\emph{after} the moment the engine emitted a sample
derived from it). Therefore:

\begin{equation}
\tilde{\ell}_I \;=\; t_e - \tilde{t}_c \;\le\; t_e - t_c^{\text{true}} \;=\; \ell_I^{\text{true}}.
\label{eq:lb}
\end{equation}

The reported $\tilde{\ell}_I$ is a verifiable \emph{lower bound} on the
true inference latency. Its slack is the engine's effective queue depth
$d$ times the frame interval ($d \times 33\,$ms at $30\,$Hz); WebGazer's
$d$ is not observable from outside the library, so we do not bound the
slack --- the number is a floor, not an estimate. We report it alongside
the FaceMesh column in \S\ref{sec:findings:results}, marked with
$\dagger$.

\subsection{Benchmark harness and analysis suite}
\label{sec:method:harness}

The harness collects samples under two protocols. \emph{Sweep} presents
targets on a full-screen $16{\times}8$ grid ($128$ cells) row-major with
a $3\,$s dwell. \emph{Drift} draws a random $N\!=\!10$ subset from a
coarser $12{\times}8$ grid ($2\,$s dwell, idle gaps), so calibration
decay shows up as a slope of per-target error against wall-clock time.
For every sample the harness records $(t_c, t_e, t_r)$ plus target and
gaze coordinates; $t_r$ is sampled in the next
\texttt{requestAnimationFrame} callback after the cursor's DOM update,
i.e.\ at hand-off to the rendering pass, before paint and composite
(\S\ref{sec:method:problem}). The CSV export carries a header block of
aggregate metrics --- including which rVFC clock served as $t_c$
(\texttt{capture\_clock\_source}) --- followed by per-sample rows.

The analysis suite computes the per-cell
mean-error heatmap, the $3{\times}3$ region partition~\cite{ivt}, the
within-fixation velocity distribution, and the linear drift slope.

\paragraph{Catching the bug in our own code.}
We hit this collapse twice. The FaceMesh engine first passed
\texttt{performance.now()} for both $t_e$ and $t_c$, reporting
$0.00\,$ms median inference latency over $7\,720$ samples; sourcing
$t_c$ from the rVFC queue (\S\ref{sec:method:exact}) raised it to
$22/27\,$ms median/$p_{95}$. The WebGazer path still read $0\,$ms after
that fix until we applied the lower-bound loop of
\S\ref{sec:method:opaque} ($34/52\,$ms). The pattern is detectable only
once the methodology is in place.

\paragraph{Reference implementation (C2).}
\label{sec:impl}\label{sec:impl:engine}\label{sec:impl:control}
We release an open TypeScript single-page application implementing the
methodology on two interchangeable engines behind a common gaze-callback
contract (Fig.~\ref{fig:arch}, appendix): the new \textbf{FaceMesh+KRR}
engine, which maps a $13$-dimensional MediaPipe-landmark feature
vector~\cite{facemesh} to screen coordinates with an RBF kernel ridge
regressor~\cite{krr}, and the \textbf{WebGazer} baseline~\cite{webgazer}
(ridge regression on eye-patch pixels), which exposes no per-frame
timestamp and so takes the lower-bound pairing of
\S\ref{sec:method:opaque}. Both feed a control layer (One-Euro
smoothing~\cite{oneeuro}, I-VT fixation classification~\cite{ivt},
dwell-clicks) and share one smooth-pursuit calibration, so comparisons
differ only in the engine. Full feature definitions, parameters, and file
paths are in the released code and Appendix~\ref{app:impl}.

% =====================================================================
\section{Empirical findings}
\label{sec:findings}

\subsection{Protocol}
\label{sec:findings:protocol}

We evaluate both engines on both tasks (sweep, drift), giving four runs,
all from a single user ($N\!=\!1$) in one session under fixed lighting,
posture, viewing distance ($60\,$cm), and window geometry: a
full-screen-width browser window of $1890\times1071$ CSS\,px on a
$14$-inch MacBook Pro (M4, 2024; $30.24\,$cm-wide panel, integrated
webcam), i.e.\ the $1512$-pt scaled desktop at $80\%$ browser zoom, so
$1890$ CSS\,px span the panel width and $1\,$cm $= 62.5$ CSS\,px.
Angular quantities use the exact conversion
$\theta(e) = \arctan\!\big(e / (62.5\,\text{px/cm} \times
60\,\text{cm})\big)$ for a pixel distance $e$
($1^\circ \approx 65.5\,$px near the screen centre); per-sample errors
in the released CSVs are in pixels, so any reader can re-derive the
angles under a different geometry. One start-of-session pursuit
calibration is shared, with no recalibration between runs. The four runs were executed in a fixed
order (WebGazer sweep, FaceMesh sweep, WebGazer drift, FaceMesh drift)
without counterbalancing, separated by short rests. The within-subject
design removes between-user variance, and the ablation
(\S\ref{sec:ablation}) surfaces a $\sim 4.6^\circ$ between-run
variability band under a filter-parameter sweep --- an imperfect
proxy for a replicate-based noise floor (\S\ref{sec:ablation}) that we
use only conservatively, to refrain from ranking. We read the
differences below against that band.

\subsection{Finding 1: a $\boldsymbol{20}$--$\boldsymbol{50\,}$ms latency gap between honest and naive measurement}
\label{sec:findings:results}

Table~\ref{tab:eval} reports the headline metrics under the capture-clock
methodology of \S\ref{sec:method}; the inference-latency columns are the
central finding. Under exact pairing FaceMesh+KRR reports
$22.0$--$22.8\,$ms median ($26.8$--$27.0\,$ms $p_{95}$); under lower-bound
pairing WebGazer reports $32.8$--$34.0\,$ms median
($50.6$--$52.0\,$ms $p_{95}$). A naive implementation reports all of
these as $\approx 0\,$ms, a $20$--$50\,$ms gap large enough to decide
whether a $50\,$ms interactive-latency target is met. FaceMesh's $p_{95}$
pipeline latency ($27$--$28\,$ms) clears it on $30\,$Hz video; WebGazer's
lower bound already exceeds the budget at $51$--$52\,$ms $p_{95}$, so its
true latency fails it by at least that margin.

\paragraph{Accuracy.}
FaceMesh's mean error is $3.1^\circ$ (sweep) and $4.6^\circ$ (drift)
lower than WebGazer's. Both differences sit inside the $\sim 4.6^\circ$
between-run variability band the ablation surfaces (\S\ref{sec:ablation}),
so at $N\!=\!1$ they are a property of this session. The grid-resolution
study (\S\ref{sec:scaling}, appendix) reinforces this: the apparent
ordering \emph{reverses} when each engine uses its own default
calibration, so the winner depends on the calibration each engine
happened to use. Neither engine reaches sub-$2^\circ$ accuracy at full
grid density, and both behave as region-prompting signals.

\begin{table}[t]
\centering
\caption{Cross-engine evaluation, all four runs ($N\!=\!1$); bold marks
the better value per task, as a reading aid only --- accuracy
differences fall within the between-run variability band
(\S\ref{sec:findings:protocol}) and we do not rank the engines.
$\dagger$\,WebGazer latency uses the
lower-bound capture clock (\S\ref{sec:method:harness}). Region columns are
mean error in the central\,/\,corner thirds of a $3{\times}3$ partition;
a dash marks buckets unsampled in drift's random subset.}
\label{tab:eval}
\small
\setlength{\tabcolsep}{4pt}
\resizebox{\linewidth}{!}{%
\begin{tabular}{llrrrrrrrr}
\toprule
Engine        & Task   & Mean$^\circ$    & Med$^\circ$     & Hit\,\%         & Cen.\,/\,Cor.$^\circ$ & Jit.$^\circ$ & $v_{p99}$$^\circ$/s & Inf.\,$p95$\,ms     & Pipe.\,$p95$\,ms    \\
\midrule
WebGazer      & sweep  &        11.11    &        11.21    &         1.56    &       8.59 / 11.94    &        2.24  &              41.8   &        52.0$^\dagger$ &      52.2$^\dagger$ \\
FaceMesh+KRR  & sweep  & \textbf{ 8.05}  & \textbf{ 7.88}  & \textbf{ 2.34}  & \textbf{ 4.42 /  9.68}& \textbf{2.49}&      \textbf{65.8}  & \textbf{27.0}       & \textbf{27.3}       \\
\midrule
WebGazer      & drift  &        11.09    &        10.56    &         0.00    &       --\,/\,11.51     &        1.83  &              40.7   &        50.6$^\dagger$ &      51.0$^\dagger$ \\
FaceMesh+KRR  & drift  & \textbf{ 6.50}  & \textbf{ 5.74}  & \textbf{20.00}  & \textbf{ 4.71 /  8.87}& \textbf{4.33}&     \textbf{145.7}  & \textbf{26.8}       & \textbf{27.4}       \\
\bottomrule
\end{tabular}}
\end{table}

Hit rate (per-cell centroid-in-cell criterion,
Appendix~\ref{app:impl}) is low for both on the sweep task, where a
$16{\times}8$ grid demands landing in a
${\sim}1.8^\circ{\times}2.0^\circ$ cell; on the drift
task FaceMesh reaches $20.0\%$ while WebGazer stays at $0\%$. Calibration
drift is not measurable over our $4.5\,$min sessions (both show a slight
negative error-vs-time slope, consistent with posture settling rather
than model decay).

\subsection{Findings 2--3: precision and per-cell error structure}
\label{sec:findings:tradeoff}
\label{sec:findings:spatial}

Spatial spread and within-fixation jitter come apart. The engines are
indistinguishable in spatial spread (radial $p_{95}$
$6.13^\circ$ vs.~$6.21^\circ$), yet within-fixation $v_{p99}$ differs
$1.6$--$3.5\times$, and the per-cell error structure differs
qualitatively (FaceMesh radial, WebGazer diagonal). Both differences fall
within the same $\sim 4.6^\circ$ variability band as the accuracy figures
above, with the full analysis in Appendix~\ref{app:findings}.

% =====================================================================
% =====================================================================
% =====================================================================
\section{Downstream utility: gaze-prompted polyp segmentation}
\label{sec:downstream}

The sections so far measure the gaze signal in isolation; we now ask
whether commodity webcam gaze is \emph{useful} as a weak supervision
label, or a lab eye tracker remains a hard requirement. We test this
directly against GazeMedSeg~\cite{gazemedseg}, the MICCAI~2024 work that
released expert EyeLink gaze annotations for medical-image segmentation
and a weakly-supervised pipeline that turns them into masks.

\subsection{Protocol}
\label{sec:downstream:protocol}

GazeMedSeg's pipeline consumes per-image fixations
$(x,y,\text{duration})$, convolves them into a gaze heatmap (isotropic
Gaussian, $\sigma{=}70$\,px), applies hierarchical thresholds and a dense CRF
to derive pseudo-masks, and trains a 2-level ensemble of from-scratch 2D
U-Nets ($224^2$) supervised only by those masks; Dice is evaluated against
ground truth on a held-out test set. On Kvasir-SEG~\cite{kvasir} (polyp
endoscopy; $900$ train, $100$ test) their expert gaze, collected on an
SR~Research EyeLink~1000 ($1000$\,Hz, $\le 0.5^\circ$), reaches $77.80$ Dice,
$94.8\%$ of the $82.12$ full-mask upper bound and ahead of bounding-box
($73.33$) and point ($73.05$) labels. Polyps vary widely in size and position
(Fig.~\ref{fig:kvasir-dataset}, appendix), which a localisation-only weak
label must capture.

We hold the downstream pipeline fixed and swap the gaze CSV it consumes.
A single non-expert annotator free-views each image (${\sim}6$\,s) on the
same laptop's webcam, and the online I-VT classifier
(\S\ref{sec:impl:control}) emits one fixation per episode in GazeMedSeg's
exact CSV schema. The screen shows \emph{no gaze cursor or feedback},
preventing the viewer from chasing a displayed estimate. That CSV replaces
the EyeLink CSV with every downstream stage unchanged. The swap
necessarily changes more than the tracker: GazeMedSeg's gaze was collected
from annotators experienced with medical images under a
locate-then-scan instruction, whereas our arm is a non-expert free-viewing
with webcam-grade calibration and online I-VT fixation extraction. The
comparison is therefore \emph{expert EyeLink collection} vs.\ \emph{our
non-expert webcam collection} through an identical training pipeline, not
an isolated hardware manipulation; \S\ref{sec:downstream:results} reads
the result accordingly.

We feed each gaze source through the unchanged pipeline and report the
downstream test Dice alongside the weak-label quality that drives it; our
$4$--$7^\circ$ accuracy (\S\ref{sec:scaling}) already predicts a poor
weak label, and the experiment quantifies how poor, and why.

\subsection{Results: webcam gaze fails as a weak label}
\label{sec:downstream:results}

We ran the full GazeMedSeg pipeline on an RTX~3090, training the $2$-level
U-Net separately on each gaze source's pseudo-masks ($15$k iterations,
$bs{=}4$, identical otherwise). Swapping the gaze CSV flips
the outcome (Table~\ref{tab:downstream}): expert EyeLink gaze yields a
usable segmenter (test Dice $0.679$) while our commodity webcam gaze does
\emph{not} ($\approx 0.000$ at every checkpoint). Training loss was
near-identical in both runs, so the webcam model fits its pseudo-masks; it
simply learns the \emph{wrong} region.

\begin{table}[htbp]
\centering
\caption{Downstream Kvasir-SEG segmentation under an \emph{identical}
weakly-supervised pipeline, swapping only the gaze source (RTX~3090,
$bs{=}4$, $15$k iters). Paper rows are GazeMedSeg's reported $bs{=}8$
numbers~\cite{gazemedseg}.}
\label{tab:downstream}
\small
\begin{tabular}{lccc}
\toprule
Weak-label source & fix-in-polyp & pseudo-mask Dice & test Dice \\
                  & (median)     & vs GT (L1/L2)    &           \\
\midrule
EyeLink-1000 (control, $bs{=}4$) & \textbf{0.90} & \textbf{0.75 / 0.78} & \textbf{0.679} \\
Webcam (ours, $bs{=}4$)          & 0.17 & 0.12 / 0.17 & $\approx\,$0.000 \\
\midrule
\emph{EyeLink (paper, $bs{=}8$)}~\cite{gazemedseg} & -- & -- & \emph{0.778} \\
\emph{Full-mask upper bound}~\cite{gazemedseg}     & -- & -- & \emph{0.821} \\
\bottomrule
\end{tabular}
\end{table}

The collapse follows the weak-label chain. Webcam fixations land inside
the polyp only $17\%$ of the time (median) versus $90\%$ for EyeLink,
with ${\sim}7\times$ the per-image scatter, so the heatmaps hit the polyp
far less often (peak inside $32\%$ vs.\ $89\%$;
Figs.~\ref{fig:downstream}--\ref{fig:worst10}, appendix). The resulting
pseudo-masks overlap ground truth at Dice $0.12$--$0.17$ (webcam) against
$0.75$--$0.78$ (EyeLink). That training ceiling is already near-useless,
and the downstream score is \emph{exactly} zero because the sparse,
off-target level-1 foreground collapses to all-background and drags the
two-level logit average below threshold. We read this as the failure of
\emph{this} non-expert, free-viewing webcam collection under the
unchanged pipeline. Because annotator expertise and viewing instruction
change along with the tracker (\S\ref{sec:downstream:protocol}), the
observed gap is an \emph{upper bound} on the hardware-only penalty ---
what it establishes is that at $4$--$7^\circ$ error a commodity tracker
gives a coarse region prompt, well short of a lesion-level label;
isolating the pure hardware effect needs a protocol-matched collection,
which the multi-user replication will include.

\paragraph{Caveats.} We trained at $bs{=}4$ (a VRAM concession) equally
for both arms, so the within-experiment contrast is internally consistent,
though our in-house EyeLink control ($0.679$) sits below GazeMedSeg's
$bs{=}8$ result ($0.778$) and is an internal reference, not a
reproduction. The arms confound hardware with annotator expertise and
viewing instruction (\S\ref{sec:downstream:protocol}), so the gap is an
upper bound on the hardware-only penalty. $36$ of $900$ webcam train
images yielded no usable pseudo-mask and were excluded from the webcam
arm's training set (EyeLink had full coverage), which if anything favours
the webcam arm. Both arms were trained with a single seed; multi-seed
runs are part of the planned replication.

% =====================================================================
\section{Discussion}
\label{sec:discussion}

\paragraph{Where the contribution lies.}
None of the components we compose is novel in isolation. The novelty is
at the \emph{measurement} layer (\S\ref{sec:method}), and it holds even
when the engine is superseded. The main threat to the empirical findings
is $N\!=\!1$: the within-task cross-engine comparison is a strict
within-subject design defensible at this $N$, but the spatial-structure
result could reflect one user's head pose, so a multi-user replication
($N\!=\!5$ on the released SPA) is in preparation.

\paragraph{Deployment and privacy.}
Lowering the hardware barrier extends gaze-prompted segmentation toward
multi-annotator deployments, but only for coarse region-level prompting
(\S\ref{sec:downstream}); the released pipeline processes frames
in-browser and persists only $(x,y)$ coordinates, and we recommend
opt-in sessions with a visible in-use indicator.

% =====================================================================
% =====================================================================
\section{Conclusion}
\label{sec:conclusion}

We presented a way to recover browser webcam-gaze latency from a
per-frame capture clock (\textbf{C1}) --- rVFC \texttt{captureTime}
where available, with every fallback explicitly reported as a lower
bound --- with exact pairing for queue-exposing engines and a further
verifiable lower bound for opaque ones. We released an open TypeScript
implementation (\textbf{C2})
and reported the findings it makes visible (\textbf{C3}): a
$20$--$50\,$ms latency gap, a precision split,
and a clinical downstream result where, with a published
weak-supervision pipeline held fixed, expert eye-tracker gaze trains a
usable polyp segmenter (Dice $0.68$) and our non-expert webcam
collection does not ($\approx 0$) --- an upper bound on the
hardware-only penalty.
A multi-user replication and an end-to-end clinical evaluation on surgical
video are in preparation.

% =====================================================================

% =====================================================================
\bibliographystyle{splncs04}
\bibliography{gazelab}

@misc{rvfc,
  author       = {{Web Incubator Community Group}},
  title        = {{HTML\-Video\-Element.request\-Video\-Frame\-Call\-back()}
                  Specification},
  howpublished = {W3C WICG Editor's Draft},
  year         = {2024},
  note         = {Editor: T. Guilbert. \url{https://wicg.github.io/video-rvfc/}, accessed 2026},
}

@inproceedings{webgazer,
  author    = {Alexandra Papoutsaki and Patsorn Sangkloy and James Laskey
               and Nediyana Daskalova and Jeff Huang and James Hays},
  title     = {{WebGazer}: Scalable Webcam Eye Tracking Using User
               Interactions},
  booktitle = {Proceedings of the 25th International Joint Conference on
               Artificial Intelligence (IJCAI)},
  pages     = {3839--3845},
  year      = {2016},
}

@article{turkergaze,
  author    = {Pingmei Xu and Krista A. Ehinger and Yinda Zhang and
               Adam Finkelstein and Sanjeev R. Kulkarni and Jianxiong Xiao},
  title     = {{TurkerGaze}: Crowdsourcing Saliency with Webcam based
               Eye Tracking},
  journal   = {arXiv preprint arXiv:1504.06755},
  year      = {2015},
}

@inproceedings{facemesh,
  author    = {Yury Kartynnik and Artsiom Ablavatski and Ivan Grishchenko
               and Matthias Grundmann},
  title     = {Real-time Facial Surface Geometry from Monocular Video on
               Mobile {GPU}s},
  booktitle = {CVPR Workshop on Computer Vision for Augmented and Virtual
               Reality (CV4ARVR)},
  year      = {2019},
  note      = {arXiv:1907.06724},
}

@inproceedings{sam,
  author    = {Alexander Kirillov and Eric Mintun and Nikhila Ravi and
               Hanzi Mao and Chloe Rolland and Laura Gustafson and
               Tete Xiao and Spencer Whitehead and Alexander C. Berg and
               Wan-Yen Lo and Piotr Doll{\'a}r and Ross Girshick},
  title     = {Segment Anything},
  booktitle = {Proceedings of the IEEE/CVF International Conference on
               Computer Vision (ICCV)},
  pages     = {4015--4026},
  year      = {2023},
}

@inproceedings{gazesam,
  author    = {Bin Wang and Armstrong Aboah and Zheyuan Zhang and
               Hongyi Pan and Ulas Bagci},
  title     = {{GazeSAM}: Interactive Image Segmentation with Eye Gaze
               and Segment Anything Model},
  booktitle = {Proceedings of The 2nd Gaze Meets ML Workshop},
  series    = {Proceedings of Machine Learning Research},
  volume    = {226},
  pages     = {254--265},
  year      = {2024},
  publisher = {PMLR},
  note      = {Earlier arXiv preprint: arXiv:2304.13844 (4 authors, predecessor title)},
}

@inproceedings{gaze2segment,
  author    = {Naji Khosravan and Haydar Celik and Baris Turkbey and
               Ruida Cheng and Evan McCreedy and Matthew McAuliffe and
               Sandra Bednarova and Elizabeth Jones and Xinjian Chen and
               Peter L. Choyke and Bradford J. Wood and Ulas Bagci},
  title     = {{Gaze2Segment}: A Pilot Study for Integrating Eye-Tracking
               Technology into Medical Image Segmentation},
  booktitle = {Medical Computer Vision and Bayesian and Graphical Models
               for Biomedical Imaging (MCV \& BAMBI), MICCAI 2016
               International Workshops, Revised Selected Papers},
  pages     = {94--104},
  year      = {2017},
  publisher = {Springer},
  doi       = {10.1007/978-3-319-61188-4_9},
}

@inproceedings{gazemedseg,
  author    = {Yuan Zhong and Chenhui Tang and Yumeng Yang and Ruoxi Qi
               and Kang Zhou and Yuqi Gong and Pheng-Ann Heng and
               Janet H. Hsiao and Qi Dou},
  title     = {Weakly-supervised Medical Image Segmentation with Gaze
               Annotations},
  booktitle = {Medical Image Computing and Computer Assisted Intervention
               -- MICCAI 2024},
  series    = {Lecture Notes in Computer Science},
  volume    = {15003},
  pages     = {530--540},
  year      = {2024},
  publisher = {Springer Cham},
  doi       = {10.1007/978-3-031-72384-1_50},
  note      = {Contributes the GazeMedSeg dataset; arXiv:2407.07406},
}

@inproceedings{kvasir,
  author    = {Debesh Jha and Pia H. Smedsrud and Michael A. Riegler
               and P{\aa}l Halvorsen and Thomas de Lange and
               Dag Johansen and H{\aa}vard D. Johansen},
  title     = {Kvasir-{SEG}: A Segmented Polyp Dataset},
  booktitle = {MultiMedia Modeling (MMM)},
  series    = {Lecture Notes in Computer Science},
  volume    = {11962},
  pages     = {451--462},
  year      = {2020},
  publisher = {Springer},
  doi       = {10.1007/978-3-030-37734-2_37},
}

@inproceedings{oneeuro,
  author    = {G{\'e}ry Casiez and Nicolas Roussel and Daniel Vogel},
  title     = {{One-Euro} Filter: A Simple Speed-Based Low-Pass Filter
               for Noisy Input in Interactive Systems},
  booktitle = {Proceedings of the SIGCHI Conference on Human Factors in
               Computing Systems (CHI)},
  pages     = {2527--2530},
  year      = {2012},
  doi       = {10.1145/2207676.2208639},
  note      = {Original title uses the euro currency symbol as the
                ``1''-suffix; spelled-out form used here.},
}

@inproceedings{ivt,
  author    = {Dario D. Salvucci and Joseph H. Goldberg},
  title     = {Identifying Fixations and Saccades in Eye-Tracking
               Protocols},
  booktitle = {Proceedings of the Symposium on Eye Tracking Research and
               Applications (ETRA)},
  pages     = {71--78},
  year      = {2000},
  doi       = {10.1145/355017.355028},
}

@inproceedings{pursuit,
  author    = {Ken Pfeuffer and M{\'e}lodie Vidal and Jayson Turner and
               Andreas Bulling and Hans Gellersen},
  title     = {Pursuit Calibration: Making Gaze Calibration Less Tedious
               and More Flexible},
  booktitle = {Proceedings of the 26th Annual ACM Symposium on User
               Interface Software and Technology (UIST)},
  pages     = {261--270},
  year      = {2013},
  doi       = {10.1145/2501988.2501998},
}

@article{krr,
  author    = {Thomas Hofmann and Bernhard Sch{\"o}lkopf and
               Alexander J. Smola},
  title     = {Kernel Methods in Machine Learning},
  journal   = {The Annals of Statistics},
  volume    = {36},
  number    = {3},
  pages     = {1171--1220},
  year      = {2008},
  doi       = {10.1214/009053607000000677},
}

@article{cholec80,
  author  = {Andru P. Twinanda and Sherif Shehata and Didier Mutter and
             Jacques Marescaux and Michel de Mathelin and Nicolas Padoy},
  title   = {{EndoNet}: A Deep Architecture for Recognition Tasks on
             Laparoscopic Videos},
  journal = {IEEE Transactions on Medical Imaging},
  volume  = {36},
  number  = {1},
  pages   = {86--97},
  year    = {2017},
  doi     = {10.1109/TMI.2016.2593957},
  note    = {The Cholec80 dataset was introduced in this paper},
}

% =====================================================================
\appendix
\section{Supplementary material}
\label{app:supp}

This appendix collects implementation detail, the grid-resolution
scaling study, the ablation, and figures relocated from the main text
for space. The main paper stands alone; everything here is released in
the accompanying source bundle.

\subsection{Implementation details}
\label{app:impl}
The FaceMesh+KRR feature vector (\S\ref{sec:impl}) is $13$-dimensional:
horizontal and vertical iris displacement relative to the inter-corner
midpoint of each eye ($4$); inter-corner distance per eye as a
head-distance proxy ($2$); horizontal asymmetry between the eyes ($1$);
and raw iris and corner coordinates normalised to face-bounding-box
space ($6$). The kernel
ridge regressor standardises each
feature (mean-centre, divide by SD with a $2\times10^{-2}$ floor against
ill-conditioning) and solves in closed form. RBF $\gamma$ is set by the
median-pairwise-distance heuristic on the standardised calibration
features (recomputed at every fit, so it varies with the calibration
distribution; \S\ref{sec:ablation}); ridge $\lambda$ is fixed at
$10^{-3}$, chosen once during development and held constant across all
runs and kernels --- neither is cross-validated.
Calibration delay-compensates each pursuit sample by a fixed $100\,$ms
lag --- the canonical smooth-pursuit onset latency assumed by
pursuit-based calibration~\cite{pursuit} --- before pairing it with the
screen target; a sensitivity analysis over this constant is future
work.

\paragraph{Metric definitions.}
The online I-VT classifier uses a velocity threshold of
$1200\,$px/s (${\approx}18^\circ$/s under the conversion of
\S\ref{sec:findings:protocol}), declares
a saccade after $2$ consecutive above-threshold frames, and requires
$\geq 3$ samples before a fixation centroid is considered stable.
\emph{Samples retained} is the fraction of raw gaze callbacks surviving
the per-cell warm-up window plus the I-VT fixation gate.
\emph{Jit.}$^\circ$ (Table~\ref{tab:eval}) is temporal precision: for
each target, the root-sum-square of the per-axis standard deviations of
its samples, averaged over targets. \emph{Radial $p_{95}$} is spatial
precision: the $95$th percentile of sample distance from the cluster
centroid. $v_{p99}$ is the $99$th percentile of instantaneous
within-fixation gaze speed. \emph{Hit} (Table~\ref{tab:eval},
\S\ref{sec:scaling}) is the per-cell centroid criterion: a cell scores
a hit when the centroid of its dwell-window samples lands inside the
cell itself, so it depends only on pixel geometry, not on the angular
conversion.

\subsection{Webcam vs.\ EyeLink gaze localisation}
\label{app:downstream}
\begin{figure}[p]
\centering
\includegraphics[width=0.8\linewidth]{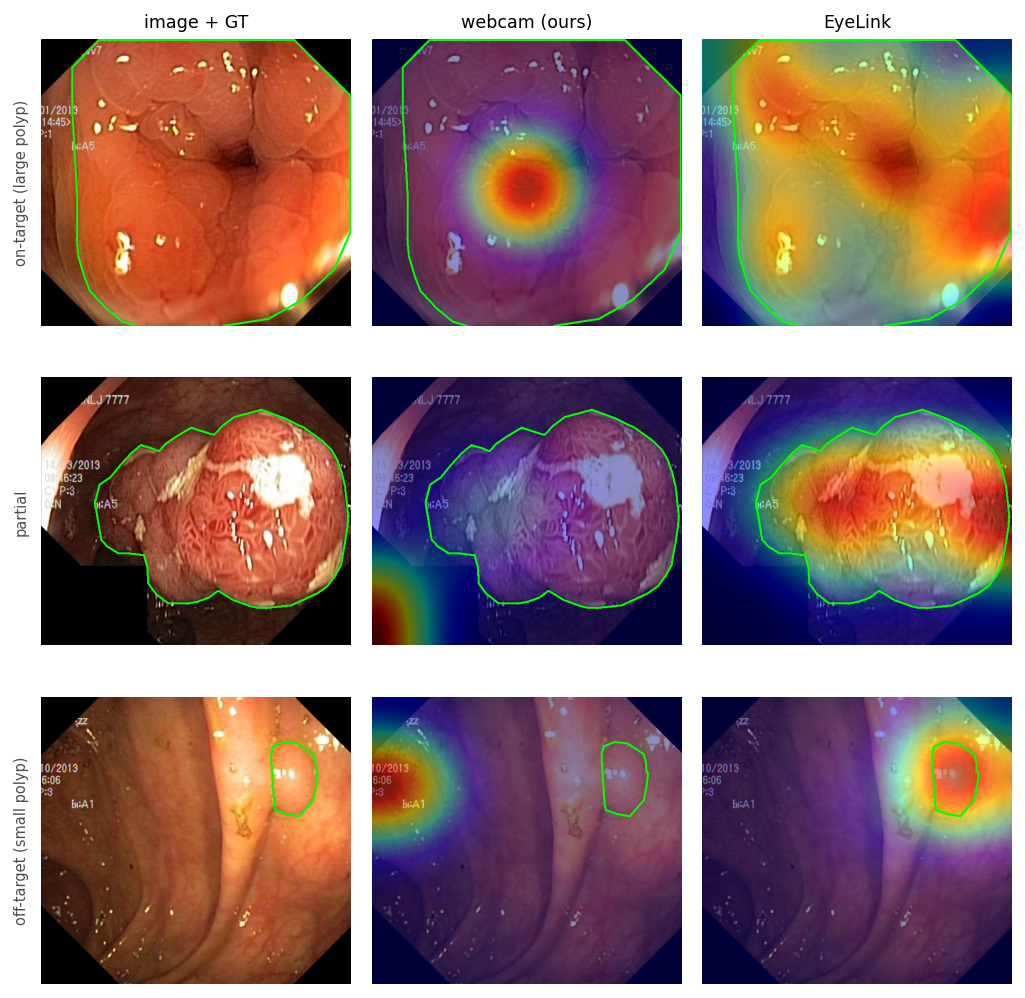}
\caption{Webcam vs.\ EyeLink gaze heatmaps on three Kvasir-SEG images
(green = GT polyp). \emph{Webcam (ours)} is our FaceMesh\,+\,KRR engine on the
laptop's integrated webcam; \emph{EyeLink} is an SR~Research EyeLink~1000
infrared lab eye-tracker, hardware orders of magnitude more expensive than a
webcam. Webcam gaze lands on large, central polyps (top) but drifts off small
or peripheral ones (bottom); EyeLink stays on-target throughout.}
\label{fig:downstream}
\end{figure}

\begin{figure}[p]
\centering
\includegraphics[width=\linewidth]{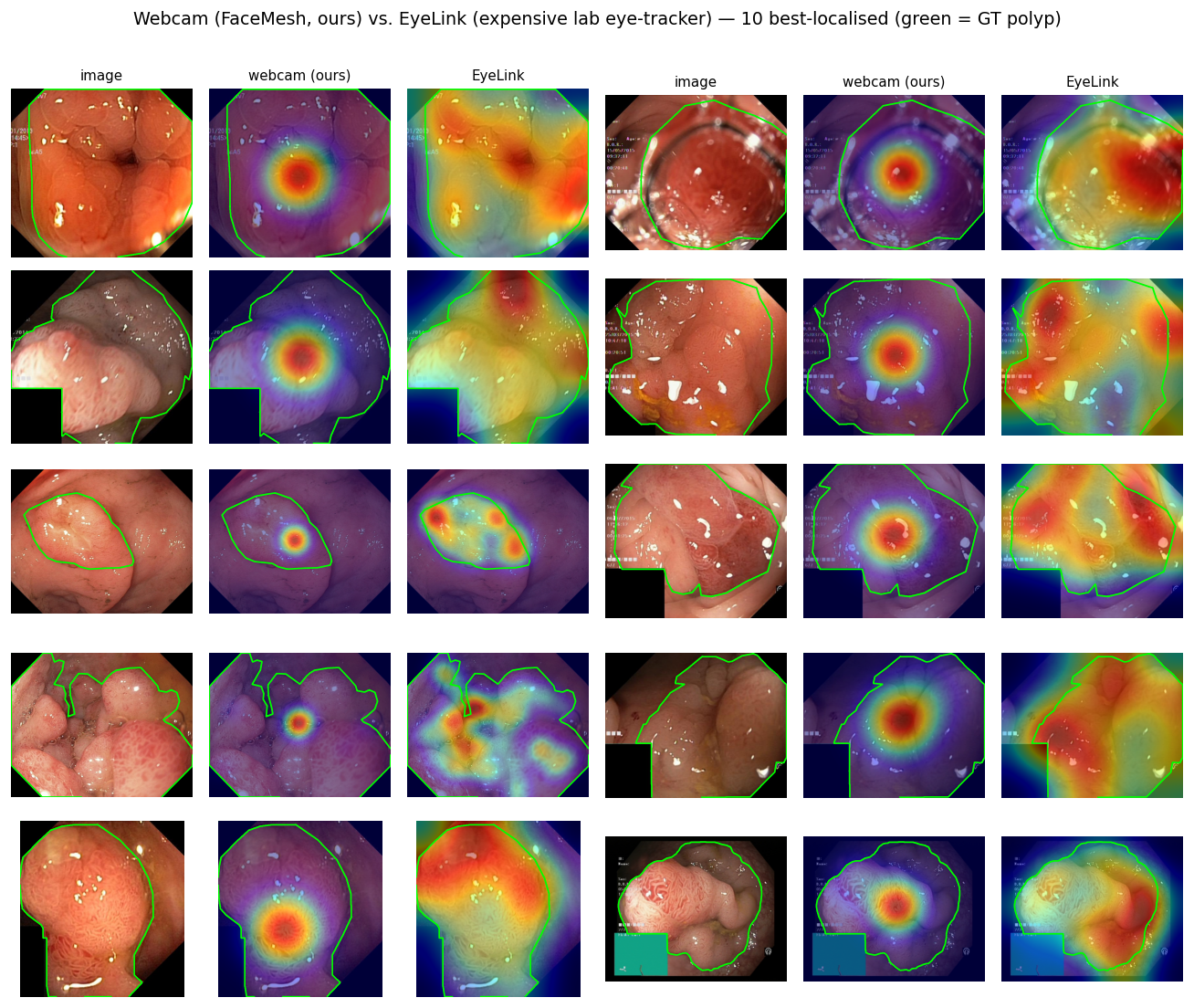}
\caption{Webcam (ours) vs.\ EyeLink gaze on the $10$ best-localised
Kvasir-SEG images; each triplet is the raw image, the webcam heatmap, and the
EyeLink heatmap (GT polyp in green). Webcam $=$ our FaceMesh\,+\,KRR engine on a
commodity webcam; EyeLink $=$ an expensive infrared lab eye-tracker.}
\label{fig:best10}
\end{figure}

\begin{figure}[p]
\centering
\includegraphics[width=\linewidth]{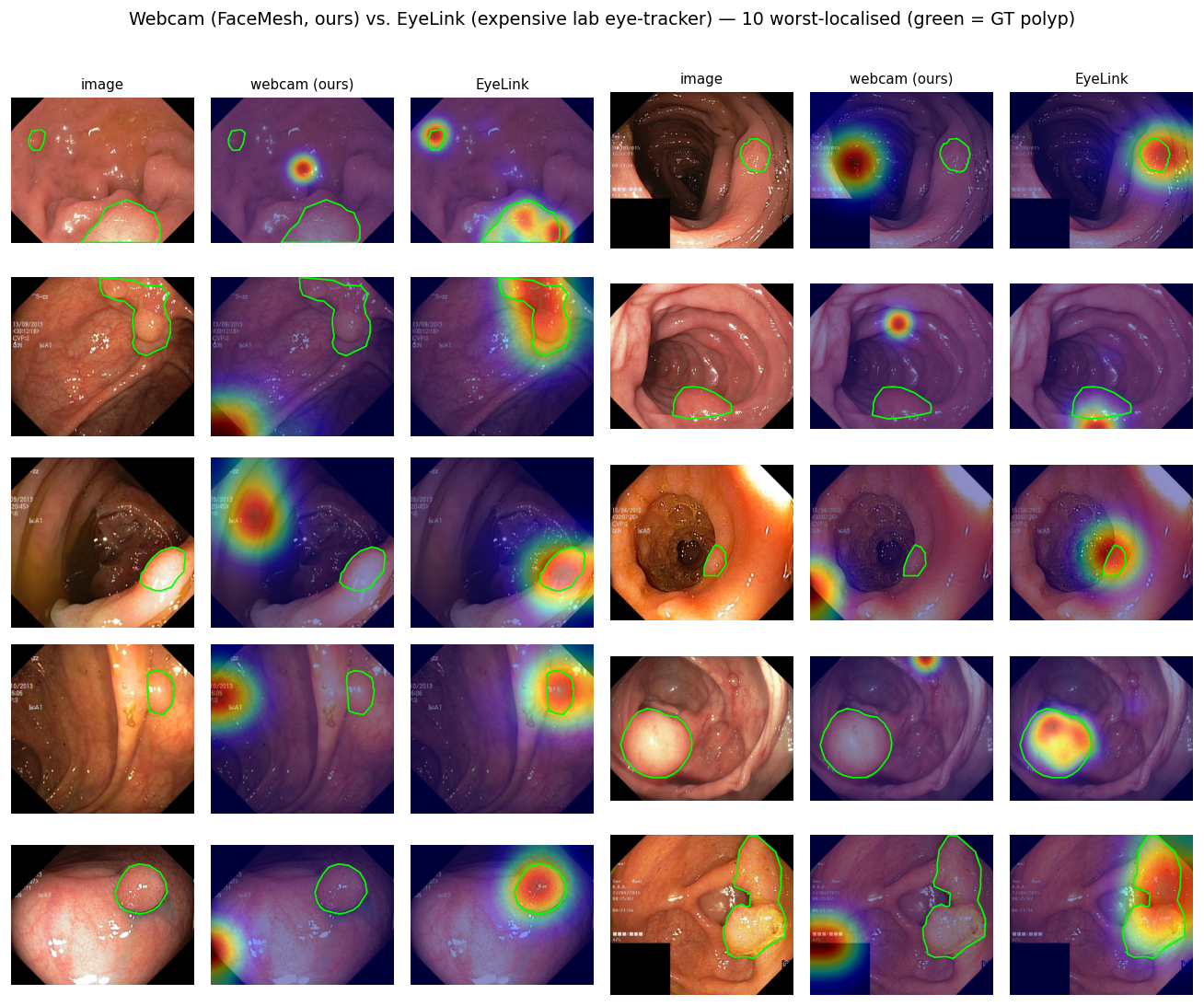}
\caption{Webcam (ours) vs.\ EyeLink gaze on the $10$ worst-localised
Kvasir-SEG images; each triplet is the raw image, the webcam heatmap, and the
EyeLink heatmap (GT polyp in green). Webcam $=$ our FaceMesh\,+\,KRR engine on a
commodity webcam; EyeLink $=$ an expensive infrared lab eye-tracker. Webcam gaze
sits near image centre or off the lesion while EyeLink remains on target.}
\label{fig:worst10}
\end{figure}

Fig.~\ref{fig:downstream} contrasts webcam and EyeLink gaze heatmaps on
three representative Kvasir-SEG images; Figs.~\ref{fig:best10}
and~\ref{fig:worst10} extend this to the $10$ best- and $10$
worst-localised images (ranked by webcam heatmap mass inside the
ground-truth polyp). On the best cases webcam gaze is visually
indistinguishable from EyeLink in landing on the lesion; on the worst it
collapses toward image centre while EyeLink stays on target, the two ends
of the $4$--$7^\circ$ accuracy floor of \S\ref{sec:scaling}.

\subsection{Kernel-ablation per-cell heatmaps}
The per-cell heatmaps for the linear and poly2 kernel ablation runs
(\S\ref{sec:ablation}) are released as PNGs in the accompanying source
bundle. The linear-kernel heatmap shows the top half of the viewport with
many cells under $1.4^\circ$ and the bottom half uniformly above
$17^\circ$, consistent with linear ridge extrapolating outside the
convex hull of the pursuit calibration trajectory. The poly2 heatmap
shows extreme outliers (one cell exceeds $59^\circ$), consistent
with under-regularised polynomial expansion at $\lambda$ tuned for
RBF.

\subsection{Extended ablation table}
The full per-run metrics, including KRR diagnostics
(calibration $N$, $\gamma$, $\lambda$, per-feature standardised
std), are provided in the accompanying source bundle.

\subsection{Reproducibility}
All raw per-sample CSV logs and the analysis scripts are provided in
the accompanying source bundle. Re-running the entire analysis from the
released CSVs takes under $30\,$s on commodity hardware; re-running the
data collection requires a webcam and a ${\sim}30$-minute time budget
for the ablation sweep (${\sim}30$-minute additional for the 4-run
paper matrix).

\paragraph{Environment.}
The runs used Chrome (arm64) on macOS Sequoia~15.6 on a $14$-inch
MacBook Pro (M4, 2024; integrated $1080$p webcam captured at
$1280{\times}720$), with MediaPipe FaceMesh \texttt{0.4.1633559619} and
WebGazer \texttt{3.4.0}; the analysis environment is
Chrome~150.0.7871.128. The collection window was a full-screen-width
browser window on the $1512$-pt scaled desktop at $80\%$ browser zoom
(\S\ref{sec:findings:protocol}); the collection-time browser build was
not logged --- the harness records the viewport, its nominal
\texttt{px\_per\_degree} assumption, and \texttt{capture\_clock\_source}
in every CSV header.
The four paper runs executed in the fixed order WebGazer sweep,
FaceMesh sweep, WebGazer drift, FaceMesh drift, with ${\geq}30$\,s rest
between runs, DevTools closed, and no window resizing; engine/task
order was not counterbalanced.

% =====================================================================
\subsection{Additional figures}
Figures relocated from the main text for space; referenced from the sections indicated in their captions.

\begin{figure}[t]
\centering
\begin{tikzpicture}[
  every node/.style={font=\footnotesize},
  block/.style={draw, rounded corners=2pt, align=center,
                inner sep=4pt, minimum width=23mm, minimum height=6mm,
                fill=white},
  layer/.style={font=\footnotesize\itshape, anchor=west, gray},
  arrow/.style={-Latex, semithick},
]
% Engine layer
\node[block]                              (cam)   {Webcam (HTMLVideoElement)};
\node[block, below=4mm of cam]            (rvfc)  {rVFC frame clock};
\node[block, below=4mm of rvfc, xshift=-16mm] (wg) {WebGazer};
\node[block, below=4mm of rvfc, xshift=16mm]  (fm) {FaceMesh\,+\,KRR};

% Control layer
\node[block, below=14mm of rvfc]          (fl)   {One-Euro filter};
\node[block, below=3mm of fl]             (ivt)  {I-VT classifier};

% Sinks
\node[block, below=4mm of ivt, xshift=-26mm] (dwell) {Dwell-click};
\node[block, below=4mm of ivt]               (csv)   {Benchmark CSV};
\node[block, below=4mm of ivt, xshift=29mm]  (heat)  {Heatmap / region};

% Arrows
\draw[arrow] (cam)  -- (rvfc);
\draw[arrow, dashed] (rvfc) -- node[left, font=\tiny, align=right]
    {lower-bound\\tag (\S\ref{sec:method:opaque})} (wg);
\draw[arrow] (rvfc) -- node[right, font=\tiny, align=left]
    {per-frame\\FIFO (\S\ref{sec:method:exact})} (fm);
\draw[arrow] (wg)   |- ($(fl.west) + (-2mm, 0)$) -- (fl.west);
\draw[arrow] (fm)   |- ($(fl.east) + (2mm, 0)$)  -- (fl.east);
\draw[arrow] (fl)   -- (ivt);
\draw[arrow] (ivt.south) |- ($(dwell.north) + (0, 2mm)$) -- (dwell.north);
\draw[arrow] (ivt)  -- (csv);
\draw[arrow] (ivt.south) |- ($(heat.north) + (0, 2mm)$) -- (heat.north);

% Layer brackets on the right, all aligned to a common x past heat.east
\coordinate (layerX) at ($(heat.east) + (6mm, 0)$);
\node[layer] at (layerX |- rvfc) {Engine};
\node[layer] at (layerX |- fl)   {Control};
\node[layer] at (layerX |- heat) {Sinks};

\end{tikzpicture}
\caption{Reference-implementation architecture. The rVFC frame
clock (\texttt{captureTime} where available, else
\texttt{presentationTime}) reaches the two engines differently:
FaceMesh+KRR receives it per frame through a FIFO queue (exact
pairing, \S\ref{sec:method:exact}), while WebGazer's samples are
tagged retrospectively with the most recent frame clock (dashed;
lower bound, \S\ref{sec:method:opaque}). The
control layer smooths raw gaze with a One-Euro filter and
segments fixation / saccade with I-VT; sinks include the
dwell-click emitter, the benchmark CSV exporter, and the
per-cell heatmap / region renderers.}
\label{fig:arch}
\end{figure}
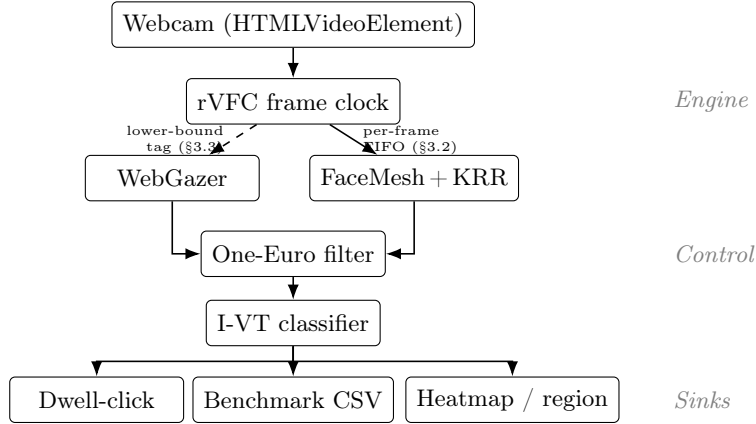

\begin{figure}[t]
\centering
\begin{tikzpicture}[
  every node/.style={font=\footnotesize},
  scale=0.9, transform shape,
]
% Eye outline
\draw[thick] (0, 0) ellipse [x radius=3cm, y radius=1.1cm];

% Corner landmarks
\fill (-3, 0) circle (2pt) node[left=2pt] {$L$};
\fill (3, 0) circle (2pt) node[right=2pt] {$R$};

% Inter-corner midpoint
\fill[gray] (0, 0) circle (1.6pt);
\node[gray, below=2pt] at (0, 0) {$m$};

% Iris (off-centre)
\fill[red!75!black] (0.9, 0.25) circle (3pt);
\node[red!75!black, above right=-2pt and 0pt] at (0.9, 0.25) {iris};

% Horizontal displacement
\draw[-Latex, blue!70!black, thick] (0, -0.6) -- (0.9, -0.6);
\node[blue!70!black, below=-1pt] at (0.45, -0.6) {$d_h$};
\draw[dashed, blue!40] (0.9, 0.25) -- (0.9, -0.6);

% Vertical displacement
\draw[-Latex, blue!70!black, thick] (-1.6, 0) -- (-1.6, 0.25);
\node[blue!70!black, left=0pt] at (-1.6, 0.12) {$d_v$};
\draw[dashed, blue!40] (0.9, 0.25) -- (-1.6, 0.25);

% Inter-corner distance
\draw[<->, gray, thick] (-3, 1.5) -- (3, 1.5);
\node[gray, above=-1pt] at (0, 1.5) {$D$ (inter-corner distance)};

\end{tikzpicture}
\caption{Per-eye landmark schematic for the FaceMesh+KRR
feature vector. From each eye we extract horizontal and
vertical iris displacement ($d_h$, $d_v$) relative to the
inter-corner midpoint $m$, and the inter-corner distance $D$
as a coarse head-distance proxy. Two eyes contribute $6$
features; a left/right asymmetry term contributes $1$
feature; the remaining $6$ features are raw iris and corner
coordinates normalised to face-bounding-box space, for a total
of $13$ dimensions per frame.}
\label{fig:features}
\end{figure}
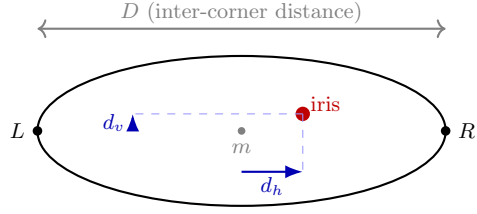

\begin{figure}[t]
\centering
\begin{subfigure}[t]{0.48\linewidth}
  \centering
  \includegraphics[width=\linewidth]{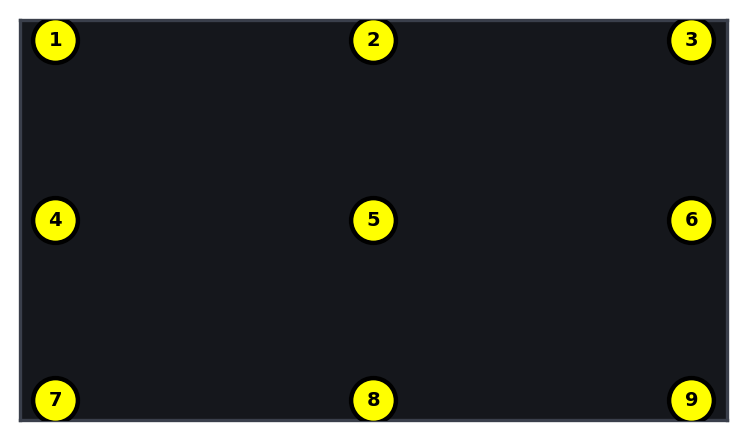}
  \caption{$9$-point click}
\end{subfigure}
\hfill
\begin{subfigure}[t]{0.48\linewidth}
  \centering
  \includegraphics[width=\linewidth]{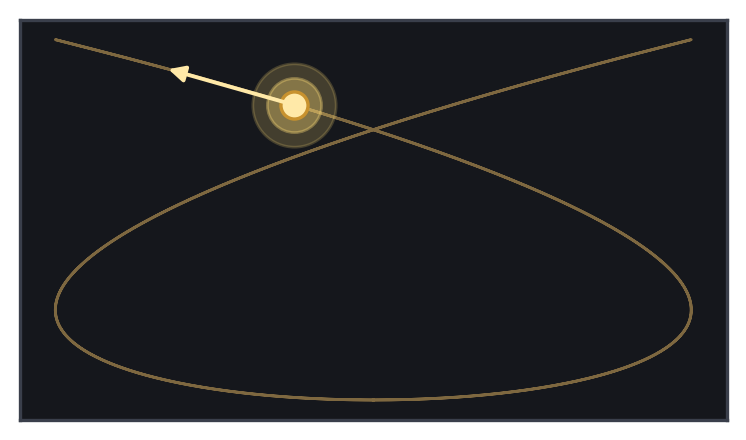}
  \caption{smooth-pursuit Lissajous}
\end{subfigure}
\caption{The two calibration target patterns, reproduced from the
reference implementation. (a) The $9$-point grid: nine fixed dots at
$\{5,50,95\}\%$ of each axis, each fixated and clicked five times
(${\sim}45$ training pairs); WebGazer's default. (b) Smooth pursuit:
a single dot traces a $3{:}2$ Lissajous curve (amplitude $0.90$ of
half-screen, so $5$--$95\%$ of each axis) for $18\,$s, yielding
${\sim}500$ pairs at $30\,$Hz; FaceMesh+KRR's default. The denser,
edge-reaching pursuit set is what lets the non-linear KRR head model
the eye-to-screen map (\S\ref{sec:impl:engine}); the two patterns cover
the same screen extent but differ by ${\sim}10\times$ in sample count.}
\label{fig:calib}
\end{figure}

\begin{figure}[t]
\centering
\includegraphics[width=\linewidth]{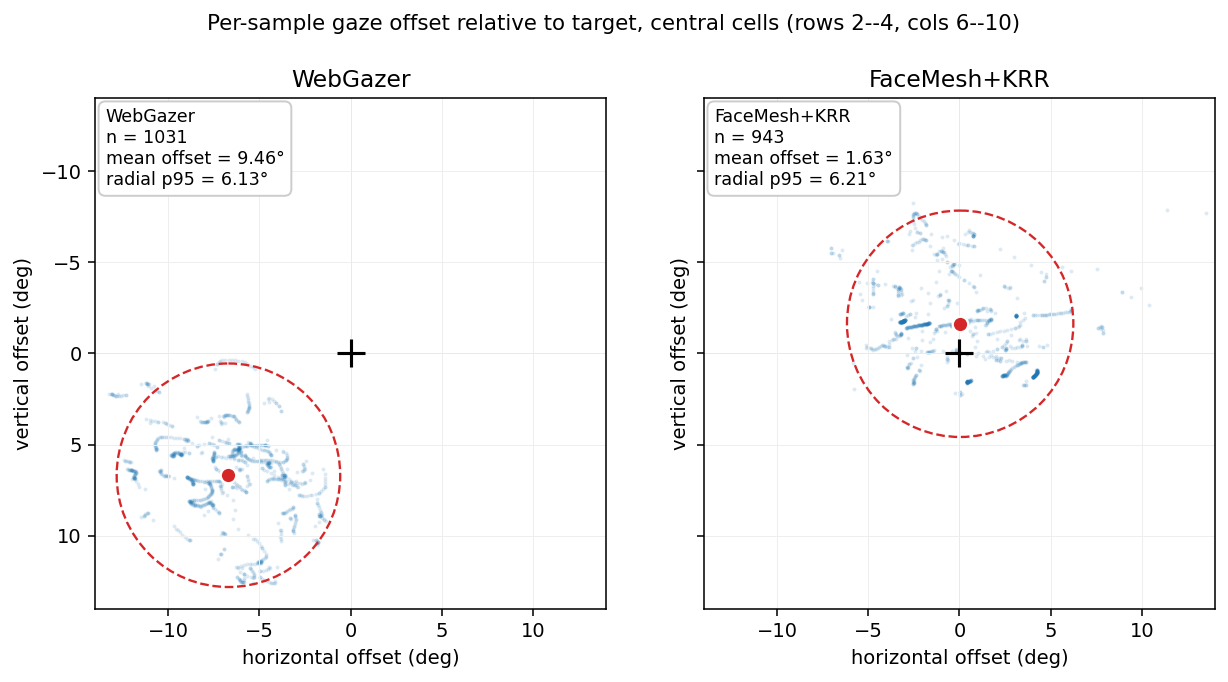}
\caption{Finding~2: per-sample gaze offset relative to target,
central cells of the sweep grid (rows $2$--$4$, columns
$6$--$10$). Each dot is one gaze sample; the black crosshair
is the target; the red dot is the cluster centroid (mean
offset, accuracy component); the dashed red circle is the
radial $p_{95}$ (spatial precision component). WebGazer's
centroid is offset by ${\sim}9.5^\circ$ into the bottom-left,
consistent with the diagonal pattern of
Fig.~\ref{fig:heatmaps}~(a). FaceMesh+KRR's centroid is within
${\sim}1.6^\circ$ of the target. The cluster spread is
comparable between engines, even though the temporal within-fixation
jitter velocity ($v_{p99}$ in Table~\ref{tab:eval}) is not:
the two notions of precision do not co-vary in this setup.}
\label{fig:scatter}
\end{figure}

\begin{figure}[t]
\centering
\begin{subfigure}[t]{0.48\linewidth}
  \centering
  \includegraphics[width=\linewidth]{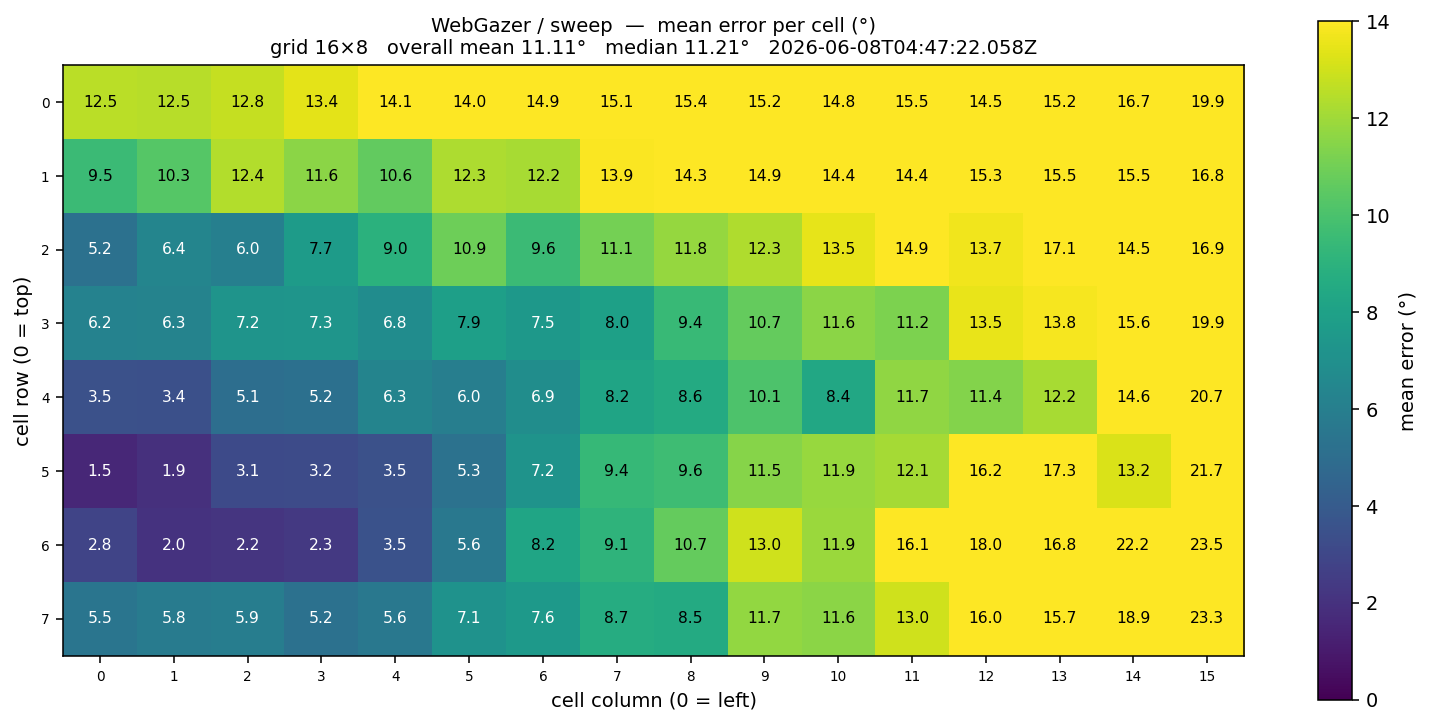}
  \caption{WebGazer / sweep ($16\times8$)}
\end{subfigure}
\hfill
\begin{subfigure}[t]{0.48\linewidth}
  \centering
  \includegraphics[width=\linewidth]{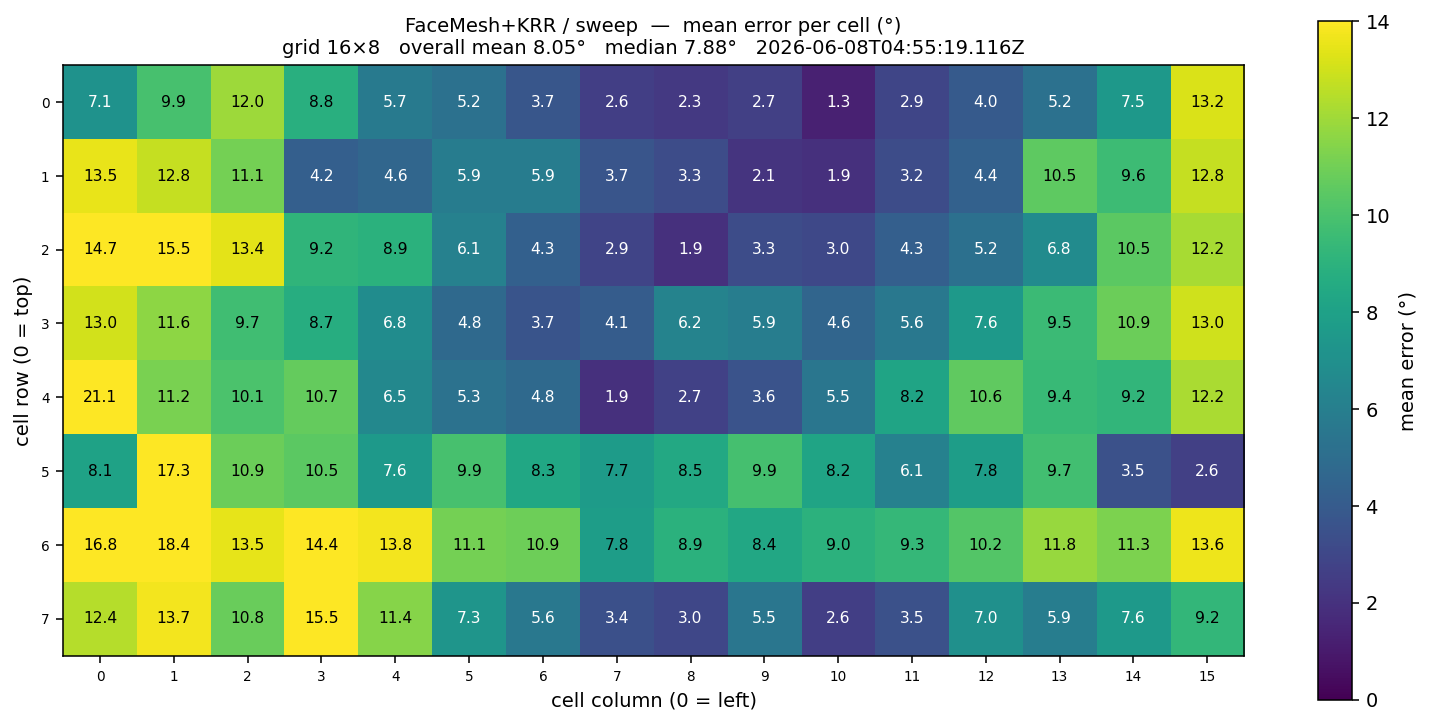}
  \caption{FaceMesh+KRR / sweep ($16\times8$)}
\end{subfigure}
\\[0.5em]
\begin{subfigure}[t]{0.48\linewidth}
  \centering
  \includegraphics[width=\linewidth]{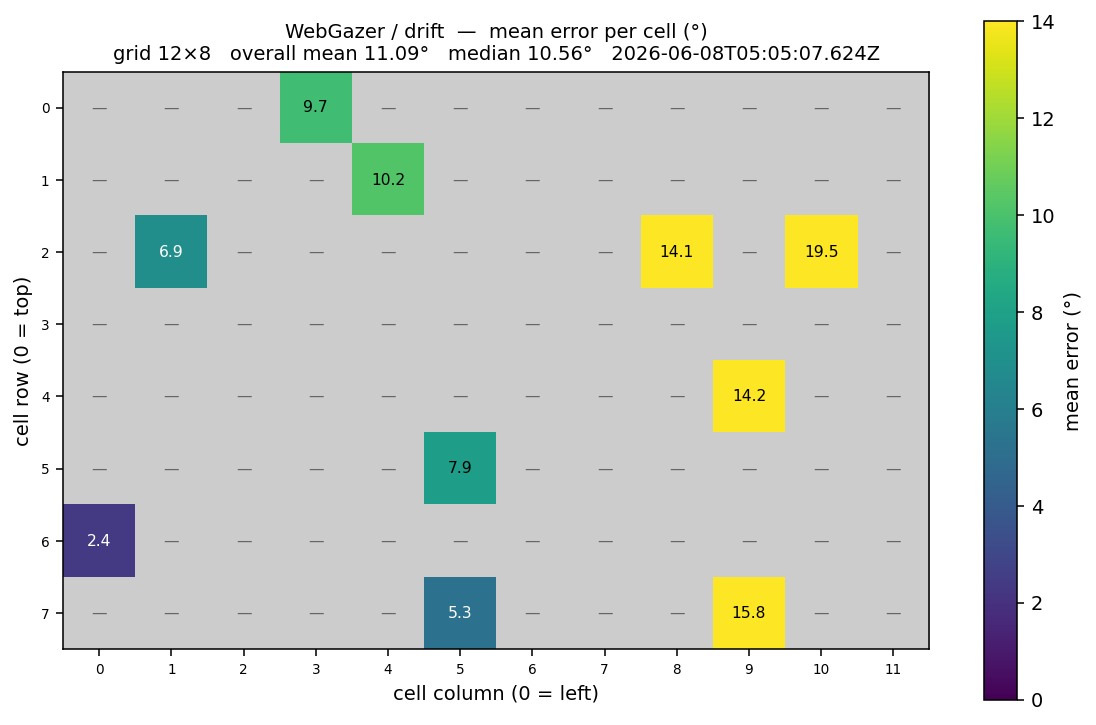}
  \caption{WebGazer / drift ($12\times8$, $10$ random visits)}
\end{subfigure}
\hfill
\begin{subfigure}[t]{0.48\linewidth}
  \centering
  \includegraphics[width=\linewidth]{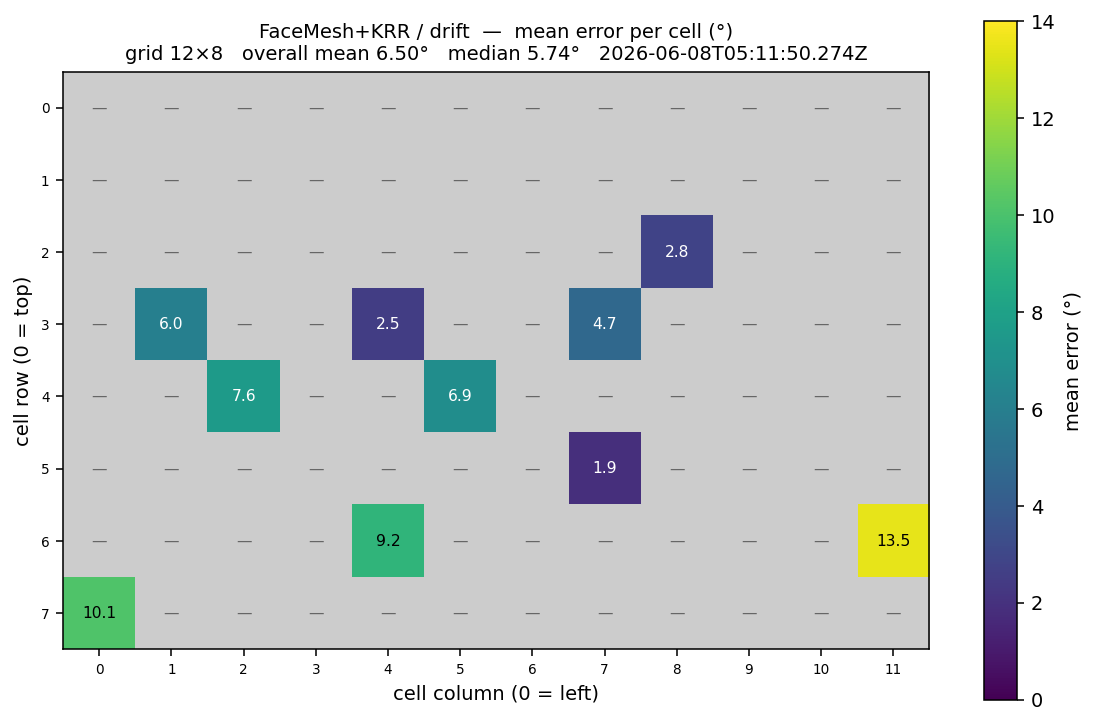}
  \caption{FaceMesh+KRR / drift ($12\times8$, $10$ random visits)}
\end{subfigure}
\caption{Finding~3: per-cell mean error ($^\circ$ of visual
angle), shared $0$--$14^\circ$ colour scale (viridis). Grey
cells were not sampled in drift's random subset. FaceMesh's
error is \emph{radially} structured around a central
low-error region; WebGazer's error is \emph{diagonally}
structured with best accuracy in the bottom-left of the
viewport. Aggregate mean / median in Table~\ref{tab:eval} does
not reveal this asymmetry.}
\label{fig:heatmaps}
\end{figure}

\begin{figure}[t]
\centering
\includegraphics[width=\linewidth]{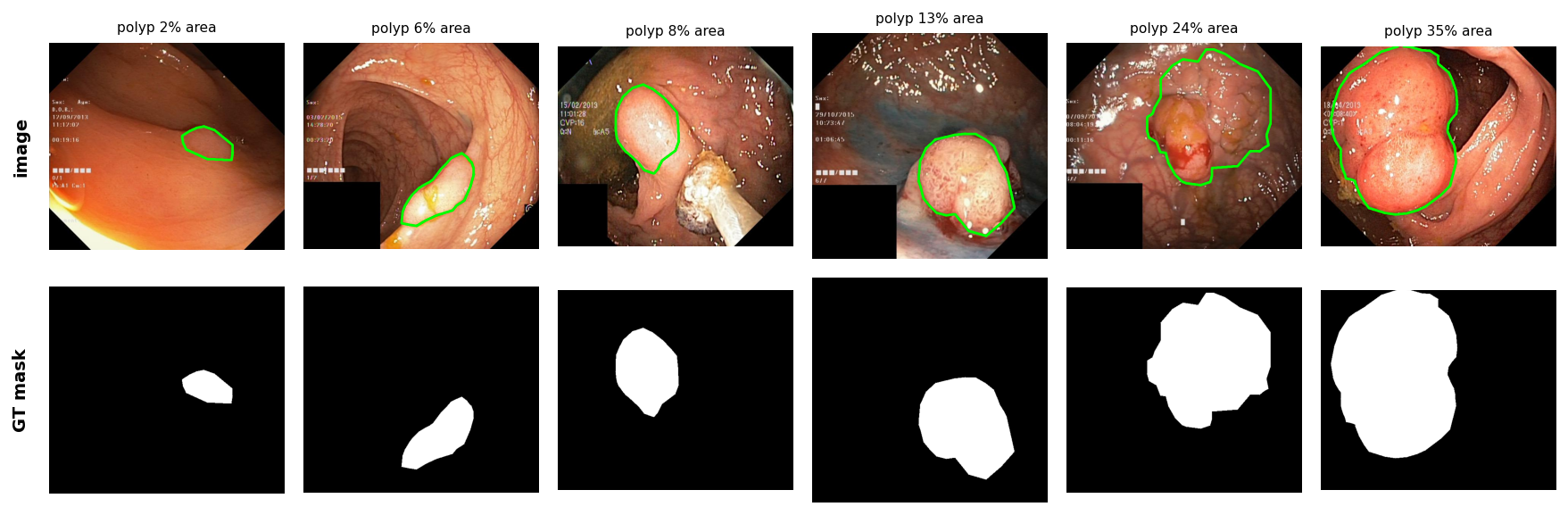}
\caption{Kvasir-SEG examples spanning the polyp-size range
(${\sim}2\%$ to ${\sim}35\%$ of image area). Top: endoscopy image with the
ground-truth polyp boundary (green). Bottom: the corresponding binary
segmentation mask. The segmentation target is the mask; gaze (expert EyeLink
or our webcam) supplies only a weak localisation prompt that the pipeline
expands into a pseudo-mask.}
\label{fig:kvasir-dataset}
\end{figure}

% =====================================================================

% =====================================================================
\subsection{Precision split and per-cell error structure (detail)}
\label{app:findings}

Spatial spread and temporal jitter do not co-vary in our data. The
per-sample offset scatter
(Fig.~\ref{fig:scatter}, appendix) places the WebGazer cluster
${\sim}9.5^\circ$ into the bottom-left and FaceMesh within
${\sim}1.6^\circ$ of the target, yet the radial $p_{95}$ spread of the
two is nearly identical ($6.13^\circ$ vs.~$6.21^\circ$), spatially
indistinguishable.
The within-fixation $v_{p99}$ column of Table~\ref{tab:eval} is the
opposite, FaceMesh $1.6$--$3.5\times$ higher, from per-frame
iris-landmark micro-movements that WebGazer's coarser features do not
expose; a pipeline can be tight on one axis and loose on the other, so
reporting one alone under-describes the engine.

The per-cell error structure also differs qualitatively
(Fig.~\ref{fig:heatmaps}, appendix): FaceMesh+KRR's error grows
\emph{radially} from a small central region (best cell $1.3^\circ$) to
the corners (worst $19.3^\circ$), as KRR extrapolates poorly outside a
pursuit trajectory that under-covers the corners, whereas WebGazer's is
\emph{diagonal} (accurate bottom-left, degraded top-right
at $17$--$23^\circ$). Both differences fall within the $\sim 4.6^\circ$
between-run variability band (\S\ref{sec:ablation}), so neither survives
as a claim about either engine until a multi-user replication.

% =====================================================================
\section{Grid-resolution scaling}
\label{sec:scaling}

The matrix of \S\ref{sec:findings} pins one grid and asks how the two
engines behave on it. This section asks the orthogonal question:
holding engine, dwell, and viewing geometry fixed, how does measured
accuracy scale as we shrink cell pitch from a trivial $1{\times}2$
partition toward a dense $8{\times}16$ grid? Separating error (does the
estimate get worse?) from per-cell hit rate (do we still land in the
right cell?) turns out to matter, because the two answer differently, and
fixes which of the two is the right lens for cross-resolution claims.

\subsection{Protocol}
\label{sec:scaling:protocol}

We sweep six grid levels, holding the aspect ratio at
rows\,:\,cols${=}1{:}2$ so each cell stays near-square on the
$16{:}9$ display and cell \emph{pitch} is the only dimension that
varies: L1 $1{\times}2$ (2 cells), L2 $2{\times}4$ (8), L3
$3{\times}6$ (18), L4 $4{\times}8$ (32), L5 $6{\times}12$ (72) and L6
$8{\times}16$ (128). We express each level by its cell pitch
$\sqrt{A_\text{screen}/N_\text{cells}}$ converted to degrees of visual
angle (\S\ref{sec:findings:protocol}), which ranges from $15.0^\circ$
(L1) down to $1.9^\circ$ (L6).
Dwell is $1.5\,$s per cell; we run four sessions
per (engine, grid) condition and interleave the two engines within
each grid so that any slow session-level drift is shared rather than
aliased onto one engine ($48$ sessions, ${\sim}1\,$h wall-clock). All
six levels share one physical setup, with the same user, seat, viewing
distance, and lighting held fixed throughout. As a fixed-grid control
on the calibration confound below, we also overlay the single-run
$8{\times}16$ baseline of \S\ref{sec:findings:results}, in which both
engines used pursuit.

Two protocol facts qualify the numbers below and we state them up
front. First, each engine uses its \emph{native default} calibration
in this sweep (pursuit for FaceMesh+KRR, nine-point for WebGazer)
rather than the matched pursuit calibration of
\S\ref{sec:findings:protocol}; FaceMesh additionally re-fits its KRR
map on every page reload, so each session carries an independent
calibration. Cross-engine comparison in this section is therefore
``engine $+$ its default calibration,'' not the engine in isolation.
Second, a posture/lighting change between the second and third repeat
of L1 depressed both engines simultaneously and persisted into L2,
inflating the L1--L2 run-to-run spread (visible as the large error
bars at the two coarsest pitches in Fig.~\ref{fig:grid-curve}). We
keep all four runs per condition and read the section for the
\emph{shape} of the curves, leaving the engine ordering open.

\subsection{Error versus cell pitch}
\label{sec:scaling:error}

\begin{table}[t]
\centering
\caption{Grid-resolution scaling, four runs per (engine, grid),
single user. Error is the per-run mean angular error
(mean\,$\pm$\,SD over four runs); hit is the per-cell
classification rate (\S\ref{sec:scaling:class}); Inf.\ is the
median inference latency. Each L1--L6 level uses each engine's default
calibration (FaceMesh pursuit, WebGazer nine-point), so columns compare
engine\,$+$\,calibration, not engines in isolation
(\S\ref{sec:scaling:protocol}). $\dagger$\,WebGazer latency excludes
two L4 runs degraded by transient host-CPU contention.
$\ddagger$\,Reference row: the $8{\times}16$ baseline of
\S\ref{sec:findings:results} at the \emph{same} grid but with
\emph{both} engines on pursuit ($N{=}1$); the WebGazer entry jumps from
$7.1$ to $11.1^\circ$, isolating the calibration effect at fixed grid
and engine.}
\label{tab:scaling}
\small
\setlength{\tabcolsep}{4.5pt}
\begin{tabular}{llrrrrrrr}
\toprule
& & & \multicolumn{2}{c}{Mean error\,$^\circ$} & \multicolumn{2}{c}{Hit\,\%} & \multicolumn{2}{c}{Inf.\,ms} \\
\cmidrule(lr){4-5}\cmidrule(lr){6-7}\cmidrule(lr){8-9}
Level & Grid & Pitch$^\circ$ & FM & WG & FM & WG & FM & WG \\
\midrule
L1 & $1{\times}2$  & 15.0 & 6.2\,$\pm$\,1.7 & 6.6\,$\pm$\,2.5 & 75 & 75 & 17 & 31 \\
L2 & $2{\times}4$  & 7.6  & 7.1\,$\pm$\,1.6 & \textbf{4.2\,$\pm$\,0.6} & 31 & 62 & 16 & 38 \\
L3 & $3{\times}6$  & 5.1  & 7.1\,$\pm$\,2.0 & \textbf{5.6\,$\pm$\,1.0} & 17 & 22 & 17 & 28 \\
L4 & $4{\times}8$  & 3.8  & 6.0\,$\pm$\,1.2 & \textbf{4.9\,$\pm$\,1.0} & 23 & 22 & 19 & 38$^\dagger$ \\
L5 & $6{\times}12$ & 2.6  & 6.4\,$\pm$\,1.8 & \textbf{5.5\,$\pm$\,0.9} & 8 & 10 & 20 & 38 \\
L6 & $8{\times}16$ & 1.9  & 9.4\,$\pm$\,1.4 & \textbf{7.1\,$\pm$\,1.9} & 1 & 3 & 22 & 44 \\
\midrule
\multicolumn{3}{l}{L6$^\ddagger$ pursuit ($N{=}1$)} & \textbf{8.1} & 11.1 & 2 & 2 & 22 & 34 \\
\bottomrule
\end{tabular}
\end{table}

\begin{figure}[t]
\centering
\includegraphics[width=\linewidth]{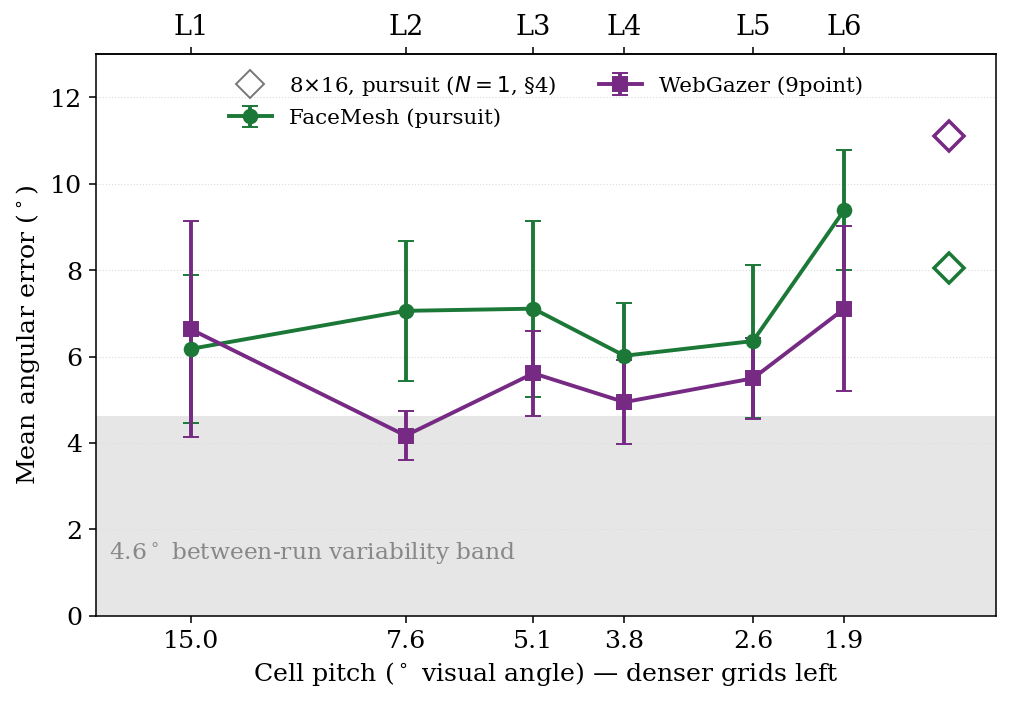}
\caption{Mean angular error versus cell pitch, one trace per engine
(mean\,$\pm$\,SD over four runs); the $x$-axis is logarithmic with
denser grids (smaller pitch) to the left, top labels mark the grid
level. Both traces are nearly flat from L1 to L5 across a ${\sim}6\times$
change in pitch ($15.0^\circ\!\to\!2.6^\circ$): error does not grow as
the grid densifies, with a modest rise at the densest grid (L6). The
shaded band is the $4.6^\circ$ between-run variability band of
\S\ref{sec:ablation}. The two open diamonds at the right (green
FaceMesh, purple WebGazer) are the $8{\times}16$ baseline of
\S\ref{sec:findings:results} ($N{=}1$) re-run with \emph{both} engines
on pursuit: WebGazer jumps from $7.1^\circ$ (its L6 square) to
$11.1^\circ$ while FaceMesh barely moves, isolating the calibration
effect at fixed grid and engine.}
\label{fig:grid-curve}
\end{figure}

Fig.~\ref{fig:grid-curve} and Table~\ref{tab:scaling} report mean
angular error as a function of cell pitch. The central observation is
that \emph{error is essentially flat against pitch}: across L1--L5 a
linear fit gives a slope of $-0.01^\circ$ per degree of pitch for
FaceMesh and $+0.10^\circ$ for WebGazer, both negligible against the
$0.6$--$2.5^\circ$ run-to-run SD. FaceMesh+KRR holds $6.0$--$7.1^\circ$
and WebGazer $4.2$--$6.6^\circ$ over that range; neither trace trends.
Densifying the grid from two targets to seventy-two does not make the
per-target estimate worse: the estimator's error is set by the
calibration and the capture geometry, not by how finely we probe the
screen. The densest grid (L6, $128$ cells) shows a modest uptick to
$9.4^\circ$ (FaceMesh) and $7.1^\circ$ (WebGazer), on the order of the
four-run SD; even there the slope stays gentle, where a
resolution-limited estimator would grow far more steeply.

The cross-engine ordering reflects the calibration method, and L6
isolates this because the same $8{\times}16$ grid was run both ways. With
each engine on its default (WebGazer nine-point, FaceMesh pursuit),
WebGazer leads at every level including L6 ($7.1^\circ$ vs.\ $9.4^\circ$).
Switching WebGazer to pursuit on that identical grid (open diamond) sends
it to $11.1^\circ$, worse than FaceMesh's $8.1^\circ$ in the
matched-pursuit regime of Table~\ref{tab:eval}. FaceMesh, always on
pursuit, barely moves. WebGazer's accuracy is far more sensitive to the
nine-point-versus-pursuit choice than FaceMesh+KRR's, so the cross-engine
gap is confounded with calibration. The L1 point carries the widest
spread for both engines, an artifact of the mid-sweep posture drift.

\subsection{Per-cell classification accuracy}
\label{sec:scaling:class}

\begin{figure}[t]
\centering
\includegraphics[width=\linewidth]{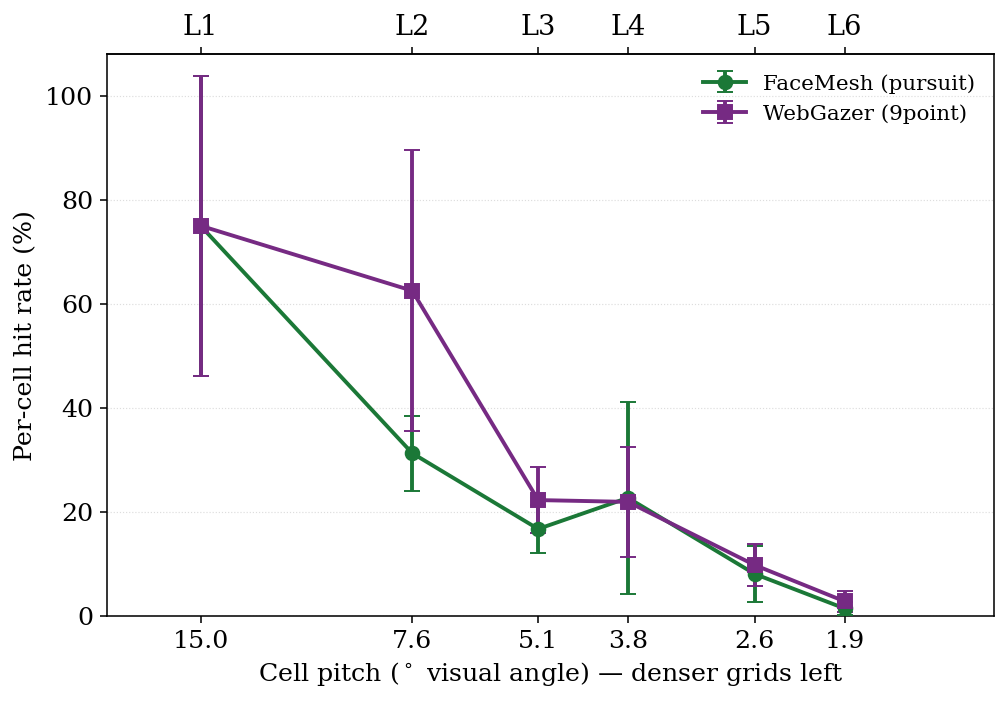}
\caption{Per-cell hit rate versus cell pitch (mean\,$\pm$\,SD over
four runs; axes as in Fig.~\ref{fig:grid-curve}). Unlike error, hit
rate falls steeply as the grid
densifies, from ${\sim}75\%$ at L1 to ${\sim}2\%$ at L6, because a
fixed gaze error clears an ever-smaller cell less often. The decline is
mechanical: tracking quality is unchanged (cf.\ the flat error in
Fig.~\ref{fig:grid-curve}). At the densest grids the hit rate and its
spread are both near zero, so the dense-grid error bars are within the
marker.}
\label{fig:grid-class}
\end{figure}

Reading each grid as a closed-set classifier (the fraction of dwell
samples whose estimate falls inside the ground-truth cell) gives the
complementary view in Fig.~\ref{fig:grid-class}. Here the curves
\emph{do} move: hit rate falls monotonically with pitch for both
engines, from ${\sim}75\%$ at L1 ($15.0^\circ$ cells) to ${\sim}2\%$
at L6 ($1.9^\circ$ cells). This fall is a direct consequence of the
flat error in Fig.~\ref{fig:grid-curve}: a ${\sim}5$--$7^\circ$
estimate lands inside a $15^\circ$ cell most of the time and inside a
$2.6^\circ$ cell almost never. Hit rate thus measures cell size against a
fixed error budget; the estimator itself is not degrading, which is why
we report cross-density accuracy through Fig.~\ref{fig:grid-curve} and
treat the classification curve as a target-size budget. With a $5$--$7^\circ$ single-session error, the
crossover below which closed-set selection stops being reliable for
this user sits between L2 and L3, i.e.\ cells no smaller than
${\sim}7^\circ$ of visual angle.

\subsection{Latency stability across grid resolutions}
\label{sec:scaling:latency}

\begin{figure}[t]
\centering
\includegraphics[width=\linewidth]{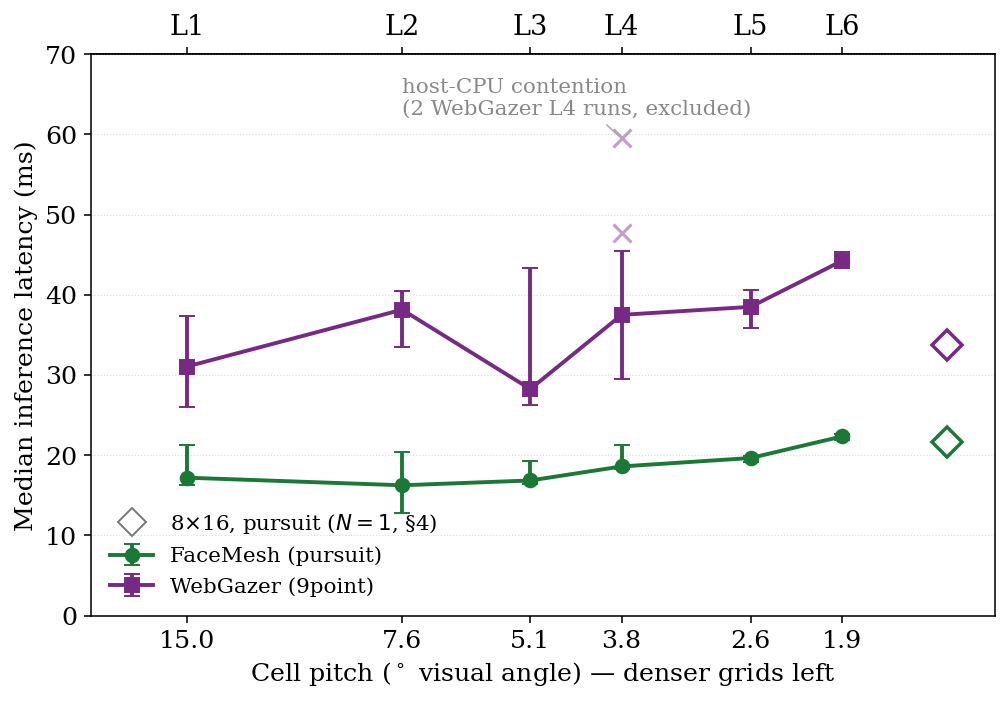}
\caption{Median inference latency versus cell pitch (markers at the
per-level median, whiskers spanning the four-run min--max; axes as in
Fig.~\ref{fig:grid-curve}). The two open diamonds at the right (green
FaceMesh, purple WebGazer) are the same pursuit $8{\times}16$ baseline
($N{=}1$) as in Fig.~\ref{fig:grid-curve}; latency, unlike accuracy, is
unchanged by the calibration switch. Both traces are flat across all
six levels: latency is set by the per-frame pipeline, not by grid
resolution. The two faint $\times$ at L4 (purple) are WebGazer runs
degraded by transient host-CPU contention ($48$ and $60\,$ms at
$24$--$25\,$Hz); they are excluded from
the median.}
\label{fig:grid-latency}
\end{figure}

The spatial protocol should not touch the capture-clock axis: cell
count changes what we draw, not the per-frame inference cost.
Fig.~\ref{fig:grid-latency} and the \mbox{Inf.\,ms} columns of
Table~\ref{tab:scaling} confirm this. FaceMesh+KRR holds a median
inference latency of $16$--$22\,$ms across all six levels (a
$64\times$ change in cell count moves it by $\le 6\,$ms), and
WebGazer holds $28$--$44\,$ms once two L4 runs degraded by transient
host-CPU contention are excluded (those read $48$ and $60\,$ms at
$24$--$25\,$Hz, with the rest of the sweep at the nominal
${\sim}30\,$Hz). L6 ($22$ and $44\,$ms) falls on the same two bands, as
does the pursuit reference ($22$ and $34\,$ms). Within each engine the
trace is flat, so the spatial and capture-clock axes are decoupled in
the harness and Fig.~\ref{fig:grid-curve} can be read at face value.
The contention episode belongs to the host machine during those two
runs, not to grid resolution; we flag it rather than smooth it.

\subsection{Where the limit appears}
\label{sec:scaling:limit}

Two limits emerge. The first is a \emph{resolution ceiling}: because
error is pitch-invariant (Fig.~\ref{fig:grid-curve}) while cell size
shrinks, usefulness for any cell-addressed sink (dwell-click
targets, region-of-interest tagging) is bounded by the ratio of the
single-session error to the cell pitch, not by the engine's behaviour
on a denser grid. For this user, with a $5$--$7^\circ$ error, that
ceiling lands around a $7^\circ$ cell (between L2 and L3); below it,
closed-set selection is noise-dominated for both engines.

The second is that most cross-condition differences do not survive the
$4.6^\circ$ between-run variability band of \S\ref{sec:ablation}. The
L3--L5 means for both engines sit within ${\sim}2.5^\circ$ of that band,
the cross-engine gaps (Table~\ref{tab:scaling}) are comparable to or
smaller than the four-run SD at the same level, and even the modest L6
uptick is on the order of that SD. The absolute numbers and the apparent
engine ordering therefore hold for a single user and session (the same
caveat as \S\ref{sec:findings:results}), and we report only the three
effects that survive the floor: within a session error is flat in pitch,
hit rate falls mechanically with it, and the cross-engine ordering is set
by calibration method (the L6 fixed-grid control). The resolution
ceiling is a practical limit the capture-clock methodology of
\S\ref{sec:method} lets us see, but the $7^\circ$ figure is this user's
alone. Iris geometry, inter-ocular distance, and habitual head pose all
feed the feature vector, so a viewer with a tighter calibration spread or
steadier posture could move the crossover a degree or two either way, and
the L1 posture episode shows how readily a single session shifts. Whether
the ceiling is a property of webcam gaze or of this one pair of eyes is
what the multi-user replication of \S\ref{sec:discussion} has to settle.
At $N\!=\!1$ we cannot.

\begin{figure}[t]
\centering
\begin{subfigure}[t]{0.32\linewidth}
  \centering
  \includegraphics[width=\linewidth]{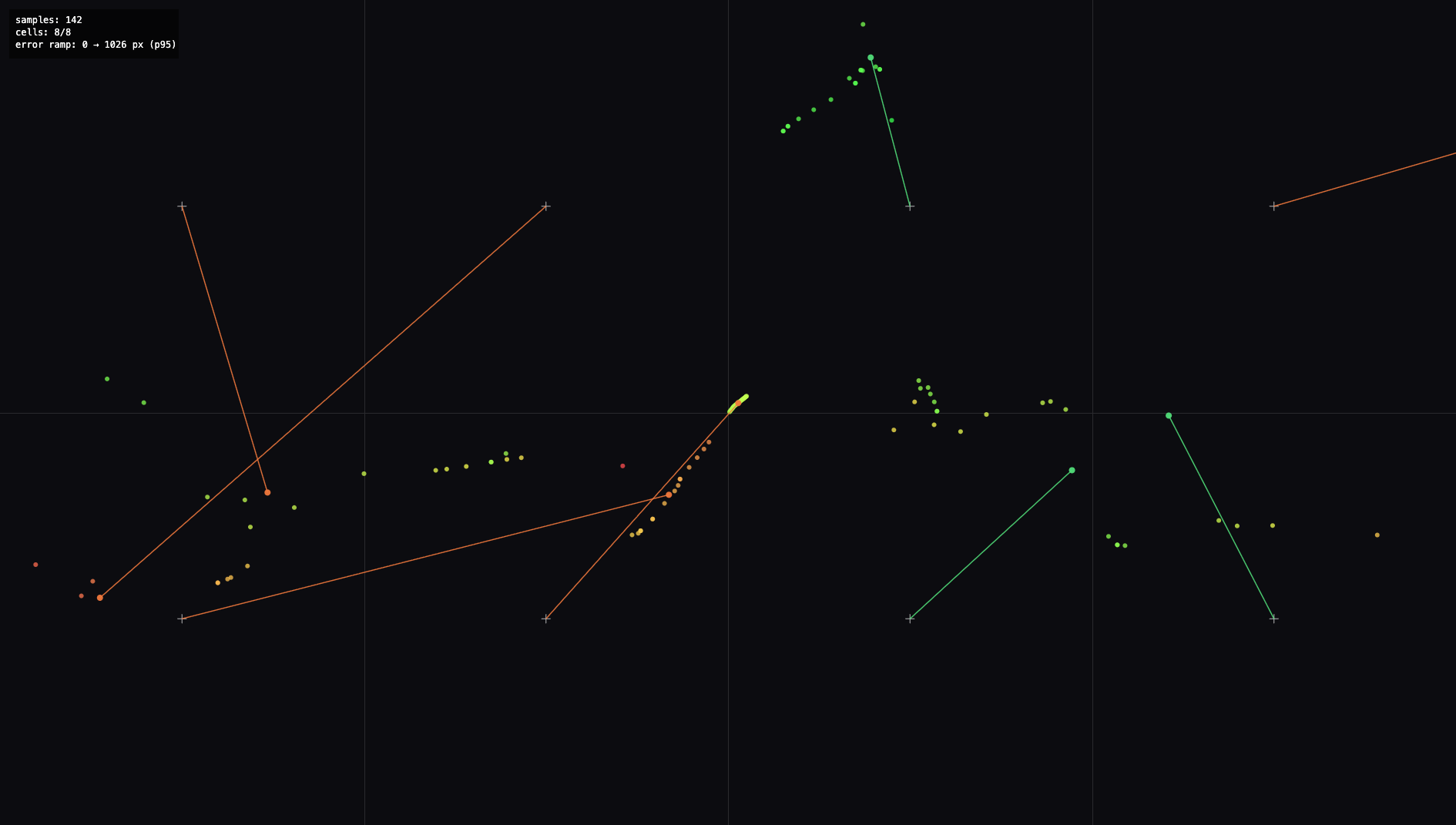}
  \caption{FaceMesh / L2 ($2{\times}4$)}
\end{subfigure}
\hfill
\begin{subfigure}[t]{0.32\linewidth}
  \centering
  \includegraphics[width=\linewidth]{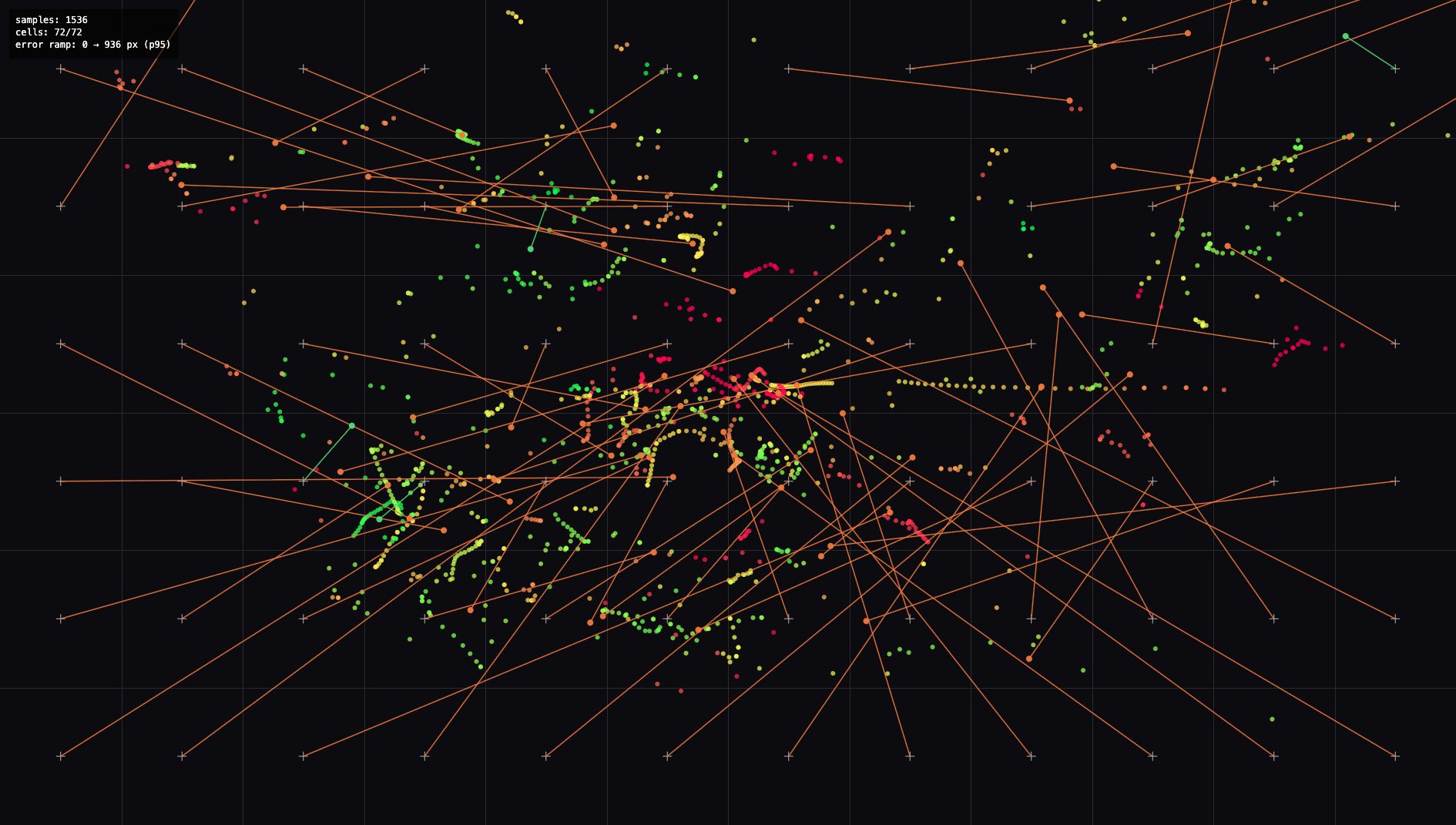}
  \caption{FaceMesh / L5 ($6{\times}12$)}
\end{subfigure}
\hfill
\begin{subfigure}[t]{0.32\linewidth}
  \centering
  \includegraphics[width=\linewidth]{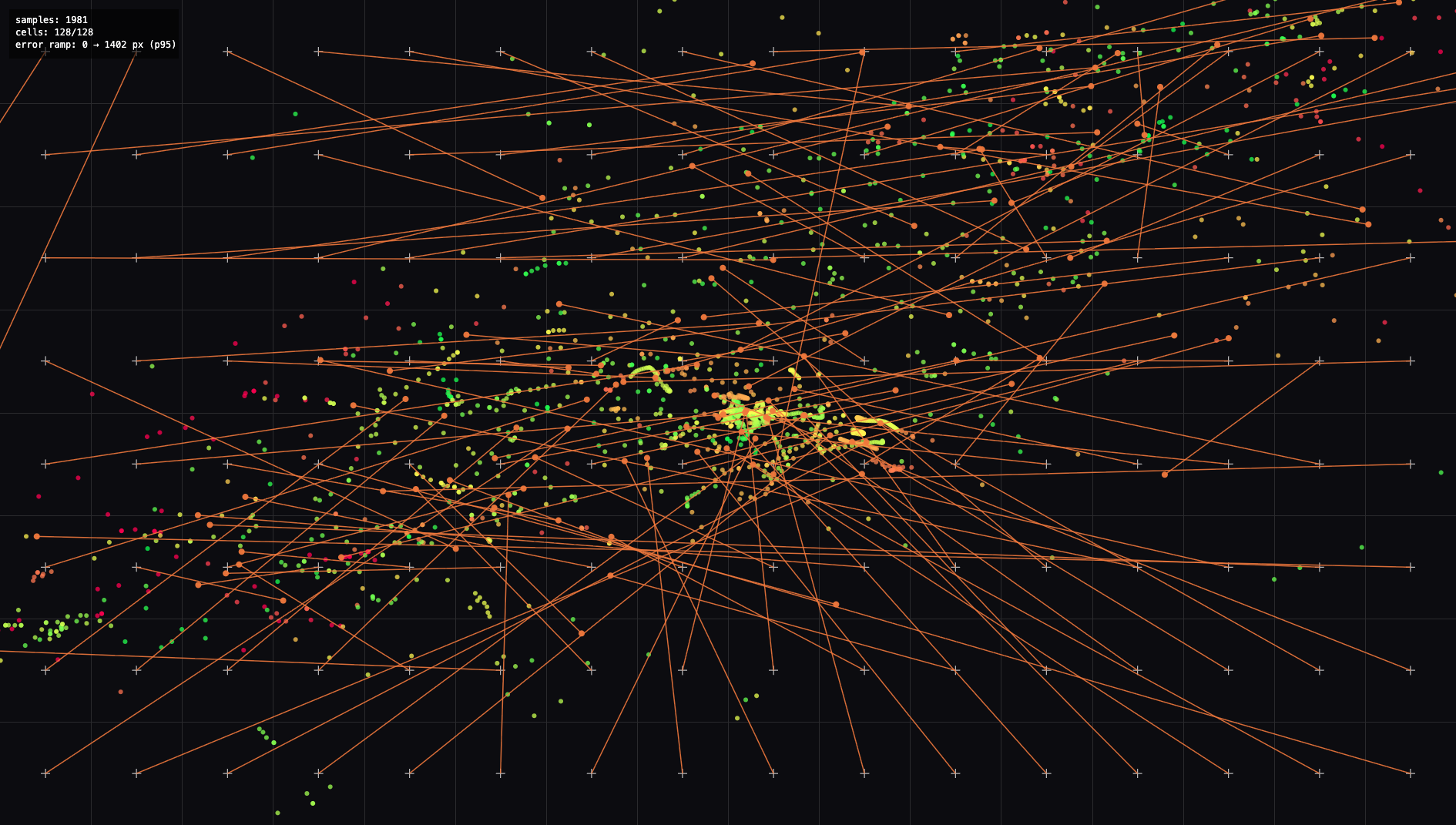}
  \caption{FaceMesh / L6 ($8{\times}16$)}
\end{subfigure}
\\[0.5em]
\begin{subfigure}[t]{0.32\linewidth}
  \centering
  \includegraphics[width=\linewidth]{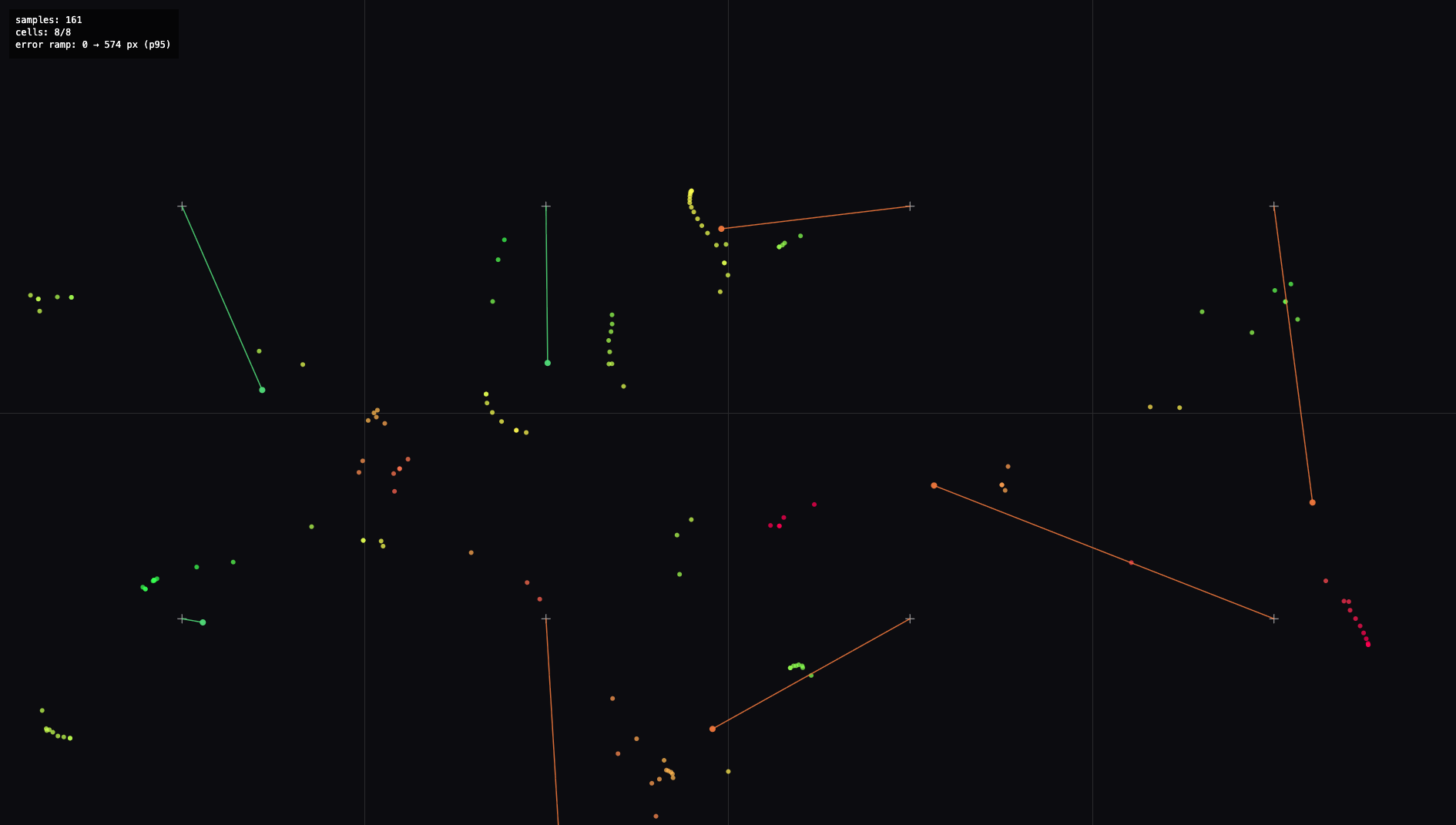}
  \caption{WebGazer / L2 ($2{\times}4$)}
\end{subfigure}
\hfill
\begin{subfigure}[t]{0.32\linewidth}
  \centering
  \includegraphics[width=\linewidth]{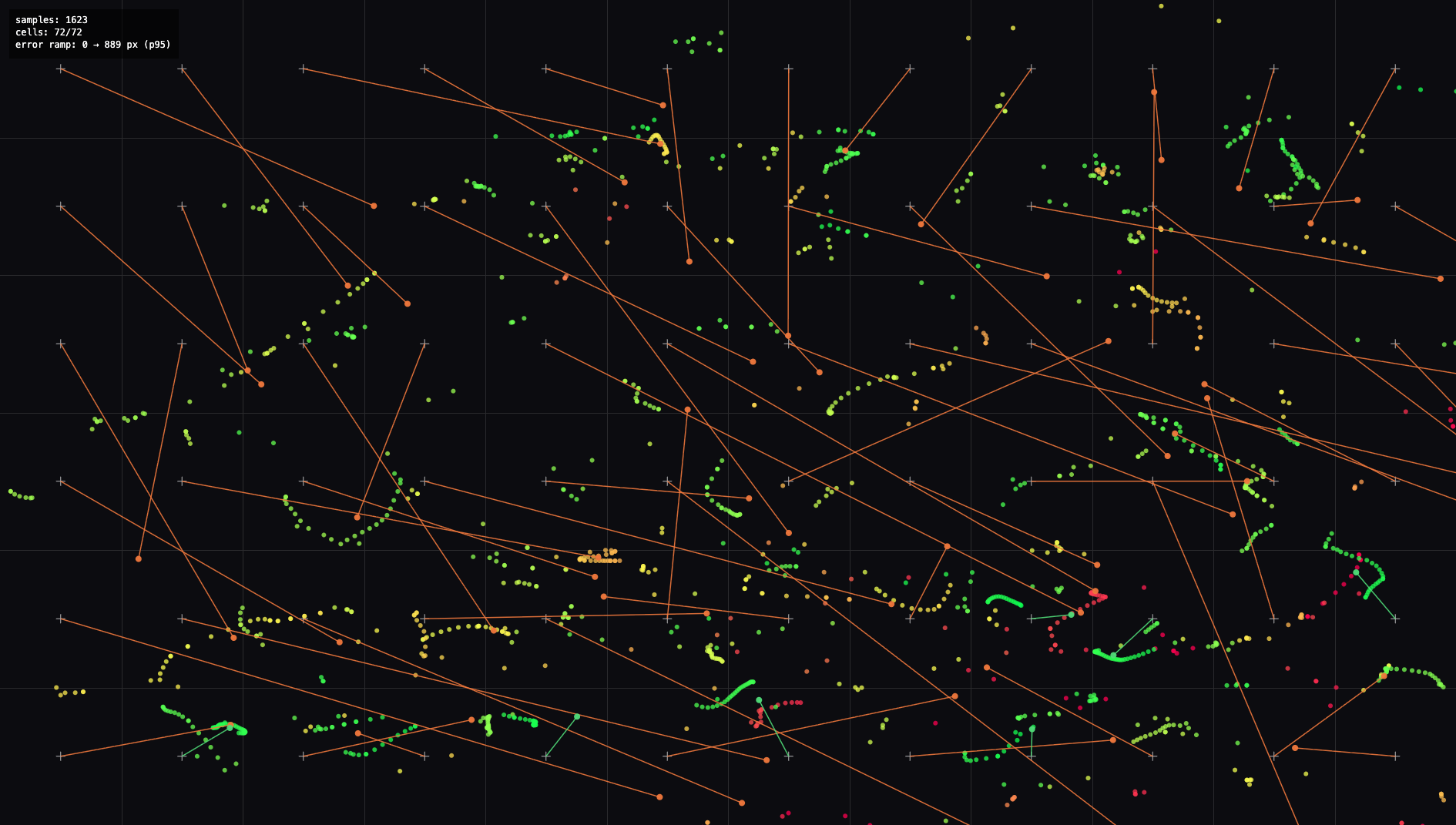}
  \caption{WebGazer / L5 ($6{\times}12$)}
\end{subfigure}
\hfill
\begin{subfigure}[t]{0.32\linewidth}
  \centering
  \includegraphics[width=\linewidth]{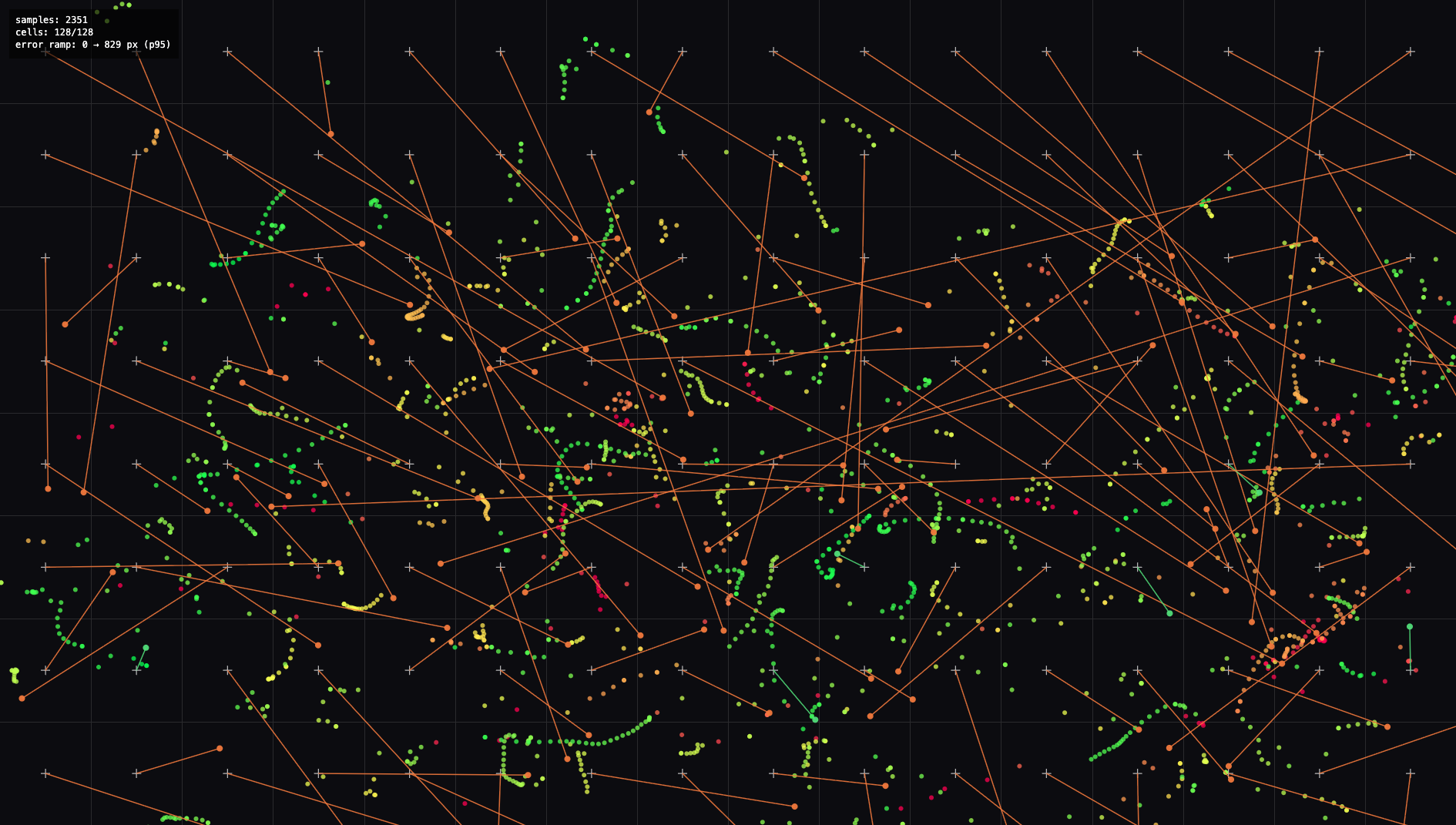}
  \caption{WebGazer / L6 ($8{\times}16$)}
\end{subfigure}
\caption{Representative gaze maps (near-mean run per condition):
dots are dwell-window samples coloured by per-sample error
(green low, red high), segments join each target to its sample
cluster. The coarse grid (left, L2) resolves into separable
per-target clusters for both engines; as the grid densifies (L5, L6)
the same ${\sim}5$--$9.5^\circ$ scatter overruns neighbouring cells,
so clusters merge: the qualitative form of the mechanical hit-rate
fall in Fig.~\ref{fig:grid-class}, while the underlying scatter is
unchanged.}
\label{fig:gazemaps}
\end{figure}

The gaze maps in Fig.~\ref{fig:gazemaps} make the ceiling concrete:
the per-target scatter is visually similar across pitches, but by L5--L6
it spans several cells, so the same estimate that classifies cleanly on
the coarse grid no longer does on the dense ones.

% =====================================================================

% =====================================================================
\section{Ablation}
\label{sec:ablation}

\begin{figure}[t]
\centering
\includegraphics[width=\linewidth]{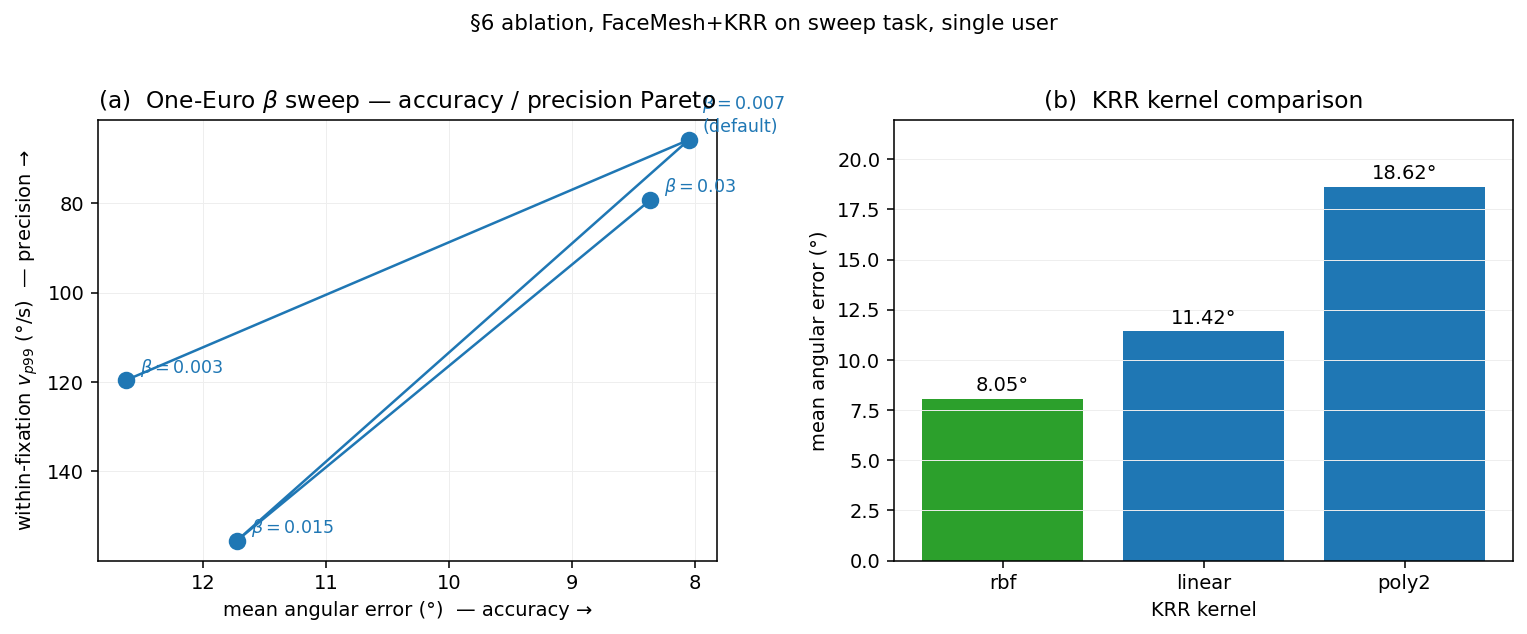}
\caption{Ablation on FaceMesh+KRR, single user. (a) One-Euro
$\beta$ sweep: each point is one full $16{\times}8$ sweep run;
$x$-axis is mean angular error (accuracy, lower-is-better,
inverted), $y$-axis is within-fixation $v_{p99}$ (precision,
lower-is-better, inverted). Points do not form a monotonic or
Pareto-shaped curve; the four conditions are noise-dominated
at $N\!=\!1$ per cell (see prose). We use the $\sim 4.6^\circ$
spread of mean error across these four runs as a conservative
between-run variability band; because the runs differ in
$\beta$ they are not replicates, so the band is used only to
refrain from ranking, not as a validated noise floor (see
prose).
(b) Three KRR kernels at default One-Euro settings. Mean error
\emph{nominally} increases from RBF through linear to poly2,
but the poly2 condition uses the $\lambda$ tuned for RBF and
is therefore under-regularised on its much higher-dimensional
feature expansion, so the bar ordering reflects regularisation, not a
clean kernel ranking.}
\label{fig:ablation}
\end{figure}

The accuracy--precision split of \S\ref{sec:findings:tradeoff} and the
KRR-vs.-ridge framing of the FaceMesh+KRR engine
(\S\ref{sec:impl:engine}) raise two natural questions that the
reference implementation can answer directly: \emph{how much of the
within-fixation jitter velocity is the One-Euro filter's responsibility
rather than the engine's}, and \emph{how much of the accuracy gain over
WebGazer comes from the non-linear kernel rather than the
$13$-dim feature space alone}. We isolate each in a one-knob sweep,
holding everything else at the protocol of
\S\ref{sec:findings:protocol}. Results are summarised in
Fig.~\ref{fig:ablation}.

\paragraph{One-Euro $\boldsymbol{\beta}$ sweep: noise-dominated.}
We re-ran the FaceMesh+KRR sweep at $\beta \in \{0.003, 0.007
\text{ (default)}, 0.015, 0.030\}$ with \texttt{minCutoff} fixed
at $1.0$.  The four-point curve does not produce a monotonic,
U-shaped, or otherwise interpretable trend:

\begin{center}\small
\begin{tabular}{lrrrr}
\toprule
$\beta$ & 0.003 & 0.007 & 0.015 & 0.030 \\
\midrule
mean $^\circ$         & 12.63 & \textbf{8.05} & 11.73 & 8.36 \\
samples retained \%   & 52    & 67             & 41    & 63    \\
within-fix $v_{p99}$ $^\circ$/s & 120 & 66 & 156 & 79 \\
\bottomrule
\end{tabular}
\end{center}

Mean angular error and samples-retained proportion are tightly
correlated across the four runs ($r \approx 0.97$): runs with
high retention are accurate, runs with low retention are not.
The calibration-time RBF $\gamma$ (set by a
median-pairwise-distance heuristic on the calibration features)
also varies across runs from $8.88\!\times\!10^{-2}$ to
$1.31\!\times\!10^{-1}$, indicating that the calibration
feature distribution itself moved between runs: user
posture and lighting drifted between runs by more than the
$\beta$ knob meaningfully altered the pipeline output.

At $N\!=\!1$ per condition the conclusion is methodological:
between-run posture and lighting variance on the order of $4.6^\circ$
swamps any signal in the tested $\beta$ range. Two caveats bound what
this spread can support. First, the four runs differ in $\beta$, so they
are treatment conditions, not replicates: the $\sim 4.6^\circ$ range can
contain genuine parameter effects on top of calibration, posture,
lighting, and sampling variation, and a true noise floor requires
repeated runs at one fixed configuration, which we have not collected.
Second, we therefore use the band in one direction only --- as grounds
to \emph{refrain} from ranking engines whose differences fall inside it
--- never to certify that an observed difference \emph{is} noise. We
recommend a $\geq 3$ replicate-per-condition protocol for future
One-Euro tuning ablations.

\paragraph{KRR kernel comparison: structural signal above the variability band.}
We compared three kernels at otherwise-default configuration:
\textbf{RBF} (the headline non-linear kernel of
\S\ref{sec:impl:engine}); \textbf{linear} (under which KRR
collapses to ridge regression on the $13$-dim feature space and
isolates the contribution of the features alone); and
\textbf{poly2} (a degree-$2$ polynomial that captures all
pairwise feature interactions without the locality property of
RBF).
Unlike the $\beta$ sweep, the kernel comparison produces a clear
structural signal (Fig.~\ref{fig:ablation}(b)):

\begin{center}\small
\begin{tabular}{lrrr}
\toprule
kernel & RBF & linear & poly2 \\
\midrule
mean $^\circ$              & \textbf{8.05} & 11.42 & 18.62 \\
hit rate (cell) \%         & 2.34          & \textbf{7.81} & 1.56  \\
samples retained \%        & 67             & 69    & 43    \\
\bottomrule
\end{tabular}
\end{center}

The mean error grows monotonically from RBF to linear to poly2,
but the \emph{shape} of each kernel's error distribution
differs qualitatively, in a way the mean alone does not show.
Linear ridge produces a \emph{bimodal} distribution
(highest hit rate of the three despite the highest mean among
RBF/linear), and the per-cell heatmap (released as
supplementary material) reveals that the bimodality is spatial:
the top half of the viewport contains many cells under
$1.4^\circ$ while the bottom half is uniformly degraded above
$17^\circ$. This is consistent with linear ridge extrapolating
beyond the convex hull of the pursuit calibration trajectory,
which under-covers the bottom edge of the screen. RBF, by
contrast, localises out-of-distribution predictions to the
training centroid (target-mean after target centring), producing
a more uniform but generally moderate error everywhere. This is the
radial structure observed in Fig.~\ref{fig:heatmaps}~(b). Poly2 is intermediate in kernel
locality but, at the same regularisation strength $\lambda
\!=\! 10^{-3}$ as RBF, produces \emph{uniformly noisy}
predictions: the diagonal-vs-off-diagonal scale of the poly2
kernel matrix is much wider than RBF's, so the same nominal
$\lambda$ provides far less effective regularisation; the
result is high-variance predictions (single cell up to
$59^\circ$ in our supplementary heatmap) that the I-VT
classifier filters aggressively (only $43\,\%$ samples
retained).

Poly2 is worse than both RBF and linear, so the comparison does not
support ``non-linear $>$ linear.'' What it does support is narrower:

\begin{quote}
\emph{The RBF kernel's locality property (predictions revert
to the training centroid out-of-distribution) produces
better full-screen behaviour than either linear ridge or
unscaled polynomial expansion at the regularisation strength
tuned for RBF. This is what produces the radial gradient
observed in \S\ref{sec:findings:spatial}.}
\end{quote}

A clean kernel comparison would re-tune $\lambda$ per kernel
(e.g.\ leave-one-out cross-validation per kernel separately).
We did not, so the poly2 column mostly shows \emph{what to control for}
in a kernel ablation.

% =====================================================================

\end{document}